\documentclass[11pt,a4,3p,times,twoside,sort,espcrc2]{elsarticle}

\usepackage{graphicx}  
\usepackage{dcolumn}   
\usepackage{bm}        
\usepackage{amssymb}   
\usepackage{textcomp}
\usepackage{mathdots}
\usepackage{booktabs}
\usepackage{color}
\newcommand{\vect}[1]{\boldsymbol{\mathbf{#1}}}

\begin{document}

\begin{frontmatter}

\title{Exact numerical methods for a many-body Wannier Stark system}

\author[unihd]{Carlos~A.~Parra-Murillo}
\author[univa]{Javier~Madro\~nero}
\author[unihd,unipa,unicl]{Sandro Wimberger,\corref{cor}}
\cortext[cor]{Corresponding Author, e-mail: s.wimberger@itp.uni-heidelberg.de, Tel.: +39 0521 90--5213}
\address[unihd]{Institut f\"ur Theoretische Physik , Universit\"at Heidelberg, 69120 Heidelberg, Germany}
\address[univa]{Departamento de F\'isica, Universidad del Valle, Cali, Colombia}
\address[unipa]{Dipartimento di Fisica e Scienze della Terra, Universit\`a di Parma, Via G. P. Usberti 7/a, 43124 Parma, Italy}
\address[unicl]{INFN, Sezione di Milano Bicocca, Gruppo Collegato di Parma, Italy}

\begin{abstract}
We present exact methods for the numerical integration of the Wannier-Stark system in a many-body scenario including
two Bloch bands. Our {\it ab initio} approaches allow for the treatment of a few-body problem with bosonic statistics and
strong interparticle interaction. The numerical implementation is based on the Lanczos algorithm for the 
diagonalization of large, but sparse symmetric Floquet matrices. We analyze the scheme efficiency 
in terms of the computational time, which is shown to scale polynomially with the size of the system. The numerically computed 
eigensystem is applied to the analysis of the Floquet Hamiltonian describing our problem.  We show that this allows, for instance, 
for the efficient detection and characterization of avoided crossings and their statistical analysis. We finally compare the efficiency 
of our Lanczos diagonalization for computing the temporal evolution of our many-body system with an explicit fourth order Runge-Kutta
integration. Both implementations heavily exploit efficient matrix-vector multiplication schemes. Our results should permit an
extrapolation of the applicability of exact methods to increasing sizes of generic many-body quantum problems with bosonic statistics.
\end{abstract}
\end{frontmatter}

\section{Introduction}
\label{sec:01}

\subsection{The Wannier-Stark problem}

The problem of a particle in a periodic potential and subjected to a constant field has a long history
in quantum physics. Today known as Wannier-Stark problem \cite{Nenciu91,Korsch}, it is the basic model for conductance. An appropriate
matching between the field and the periodic lattice can induce resonance phenomena, and hence be used for 
the control of transport. This idea goes back to Esaki and Tsu who suggested to implement diodes and other
transport devices with the techniques of semiconductor physics \cite{chan74,leo03}. While of interest
for transport, it was actually shown quite early that the constant field, also called tilt, induces regular
periodic oscillations instead of an acceleration along macroscopic sizes. This phenomenon of Bloch oscillations
and the fact that the Wannier-Stark problem is strictly speaking an open system has evoked much interest over decades \cite{Nenciu91}.

Modern realizations of the Wannier-Stark system use either light \cite{Pert99} or ultracold atoms for which the potentials
can be controlled very well \cite{ChSalomonPRL1997,Mors06}, avoiding intrinsic effects of unwanted disorder and uncontrolled particle-particle interactions in solids.
Implementations with cold atoms have turned the attention from fermions also to bosonic transport particles \cite{ChSalomonPRL1997,Mors06},
from single-particle models \cite{Korsch} over mean-field effects \cite{Mors02,thom03,Wimb05,BOweak} towards many-body correlation 
effects \cite{greiner2009,Innsbruck2013,Nagerl14,kolo03,kol68PRE2003,Tomadin,Ploetz,Carlitos01,Carlitos03}. Needless to say that the many-body scenario is much
more complicated. One typically has to resort then to effective models \cite{Sachdev,Ploetz}, which themselves build on a good intuition 
of the underlying physical phenomena that take place during the temporal evolution. The coupling between energy bands (Bloch bands in the
periodic lattice) induced by the tilt makes effective single-body descriptions already non-trivial \cite{Nenciu91,Martinelli,Korsch,SWSchlageck}, 
not to speak about the growing size of the system when including many particles in various dimensions. Here the Hilbert space grows 
exponentially, both with the size of the system and the number of particles.

\subsection{Our approach}

In view of the complications due to many coupled energy bands, spatial dimensions and many-body effects, we restrict here to a one-dimensional realization
including fully two energy bands. Our system is thus closed, but it represents a good approximation to a possible realization with ultracold atoms in a double-periodic
potential producing a miniband structure, see Fig. \ref{fig:a} below and the appendix in \cite{Carlitos01}. Here we present a detailed study of our computational methods which allow for
strong interactions between the particles. We study finite systems, with a finite number of particles, bosons to be specific, in a finite number of sufficiently
deep lattice sites. All these assumptions make it possible to use a lattice model, whereby the standard Bose Hubbard model \cite{Sachdev,Ibloch} is much 
extended here by including two coupled energy bands, interband interactions terms, and field-induced couplings. The exponential growth of the size of the
problem with, e.g. the particle number, is especially related to the choice of bosons, which can, in principle, all sit on the same lattice site. This makes our problem
non-integrable and computationally hard even in one spatial dimension \cite{kolo03,Carlitos01}. Applying periodic boundary conditions, the 
constant field is mapped onto a periodically time-dependent quasi-momentum, making our system effectively driven by a periodic force. This corresponds to an 
acceleration of the lattice as done also experimentally \cite{ChSalomonPRL1997,Mors02}. We make the problem numerically tractable by combining well-known 
methods, the Floquet theory for periodic time-dependent systems \cite{Shirley1965}, the Lanczos diagonalization algorithm 
\cite{Lanczos,ParletScott1979,GeneBook,BuchJosaB1995,Krug}, and standard propagation methods based on the Runge-Kutta scheme. For both, the diagonalization
and the propagation we heavily exploit efficient matrix-vector multiplications, storing just the non-zero matrix elements.

Our results represent benchmarks for the 
exact numerical treatment of our Wannier-Stark problem and similar many-body lattice systems with external drives. This allows for the extrapolation of
the applicability of our exact methods as a function of memory and computation time, and in view of the physical reality of finite size many-body quantum
systems. Moreover, our results could be interesting as well for interaction effects in the quantum simulation of many particle interference effects with
bosonic statistics \cite{Preiss2014}, a problem known as boson sampling \cite{Boso13}. As a paradigm application of the numerical methods for our Wannier-Stark system, we analyse its quasi-energy (or Floquet) spectrum, which is characterized by a large number of avoided level crossings in the presence of strong interactions. To do so, we present an efficient and reliable way of detecting a large number of such avoided crossings and their distributions, based on the properties of the eigenvectors. This allows us to characterize the non-integrability of our
model, and, in principle, of other many-body lattice systems in general \cite{Buon07,PloetzLubach}.

\subsection{Structure of the paper}

In Section \ref{sec:02} we introduce our many-body boson system and we briefly summarize its previously studied properties (see 
refs.~\cite{Carlitos01,Carlitos02,PhDThesis2013}). 
The main numerical tools for obtaining the eigenspectrum of our system are presented in Section~\ref{sec:03} along with a detailed analysis of their efficiency. 
Section~\ref{sec:04} is devoted to the quantum spectrum around resonance conditions, for which the two bands are strongly coupled. In particular, 
we present an alternative method for detecting and characterizing avoided level crossings. In Section~\ref{sec:05}, we extend the discussion of the spectral properties to the time domain. This turns out to be useful for the understanding of the impact of the avoided crossings and their distributions on the system's dynamics. For this, we compare the numerical efficiency of computing the time evolution using either the Floquet eigenstates and eigenspectra or a direct Runge-Kutta integration. Finally, Section~\ref{sec:06} concludes the paper with a brief discussion of the applicability of our methods to similar quantum
lattice problems.

\section{Bose-Hubbard model for the Wannier-Stark system}\label{sec:02}

The Wannier-Stark system is an interesting quantum problem because of its non-intuitive phenomenology, including Bloch oscillations (responsible for the 
effective Stark localization of wave packets). To describe interband coupling effects, such as resonant tunneling between different Bloch bands, at least two bands must be included into the model. Allowing also for many-body correlations, the problem immediately becomes very challenging \cite{kol68PRE2003,Tomadin,Ploetz,PloetzLubach,Carlitos01}.
We restrict here to a closed two-band model, which (a) could be readily implemented in experiments in good approximation, see the appendix in \cite{Carlitos01}, and (b) includes strong
interparticle interactions and interband couplings. Our many-body model for bosons is based on the celebrated Bose-Hubbard Hamiltonians, 
usually studied just within a one-band approximation \cite{Ibloch,Sachdev,kolo03,kol68PRE2003}. 

\begin{figure}[t]
\centering 
\includegraphics[width=0.8\columnwidth]{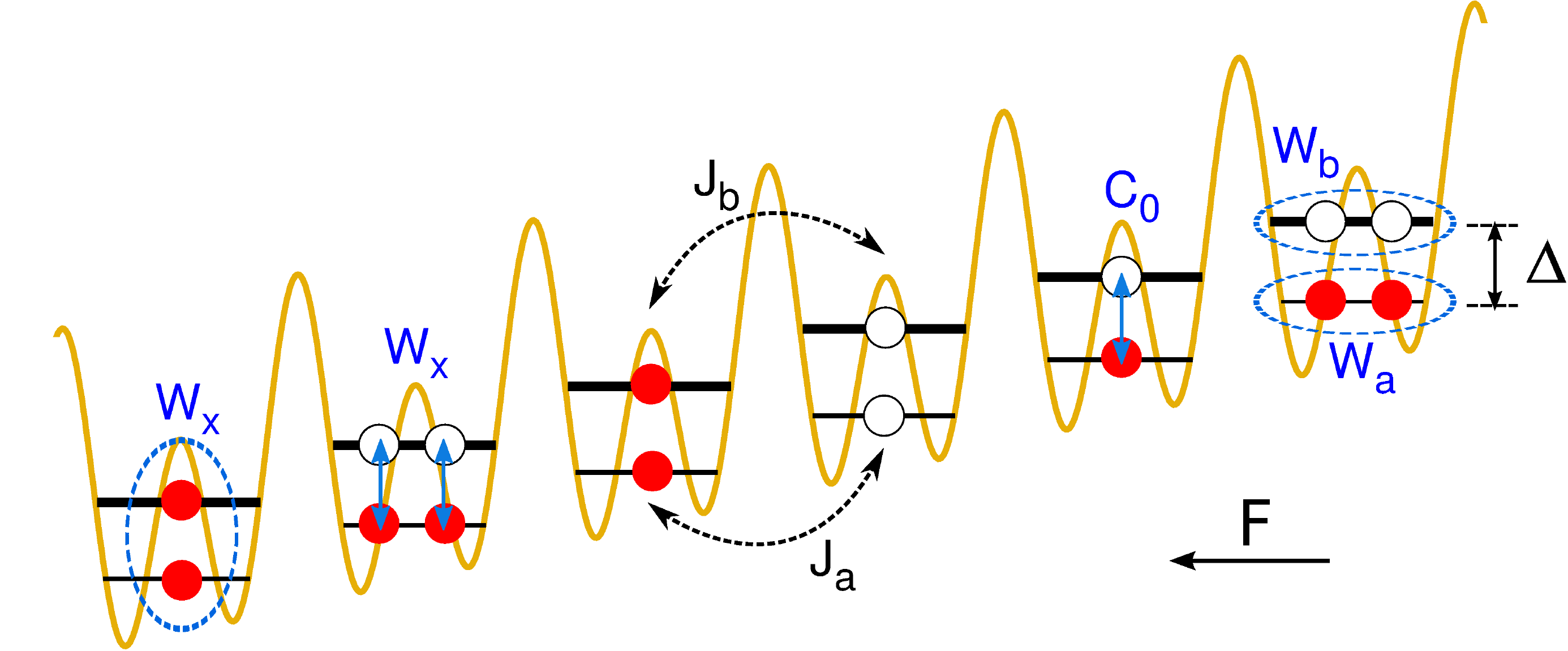}
\caption{\label{fig:a}(Color online) Sketch of the experimental realization of the system described by (\ref{eq:01}) in a double periodic potential. The considered processes of hopping and two-particle interactions are pictured by $J_a, J_b$ and $W_x, W_a, W_b$, respectively, where $a$ indexes the lower and $b$ the upper band. The Stark force $F$, introducing interband transitions, e.g. on-site with a coupling strength $C_0$, acts to the left in our sketch. $\Delta$ is the band gap between the two energy ladders corresponding to the two considered energy bands.
}
\end{figure}

Our Hamiltonian
\begin{eqnarray}\label{eq:01}
 \hat{H} &=&\sum_{l,\beta}\left[-\frac{J_{\beta}}{2}\left(\hat{\beta}^{\dagger}_{l+1}\hat{\beta}_{l}+h.c.\right)+
 \frac{W_{\beta}}{2}\hat{\beta}^{\dagger 2}_{l}\hat{\beta}^2_{l}+\omega_Bl \hat n_l^{(\beta)}\right]\,\nonumber\\
 &+&\sum_{l,\mu}\omega_B C_{\mu} \hat{a}^{\dagger}_{l+\mu}\hat{b}_l+
\frac{W_x}{2}\hat{b}^{\dagger 2}_{l}\hat{a}^2_{l}+h.c.\nonumber\\
 &+&\sum_l2W_x\hat{n}^a_l\hat{n}^b_l+\frac{\Delta}{2}(\hat{n}^{b}_l-\hat{n}^{a}_l)\,,
\end{eqnarray}
represents a two-band tight-binding or discrete-lattice model. Its possible experimental realization in a double periodic optical lattice and the considered microscopic processes are sketched in Fig. \ref{fig:a}. Our model system has the following characteristics: 

\begin{itemize}

\item[($i$)] it preserves the total number of particles (bosons) $N$ living on $L$ lattice sites or potential wells. Then the full dimension of the state (or Fock) 
space is given by
\begin{equation}\label{eq:02}
d_F=\frac{(N+2L-1)!}{N!(2L-1)!}\,.
\end{equation}

\item[$(ii)$] The operators $\hat\beta_l$ and $\hat\beta^{\dagger}_l$ represent the annihilation and creation operators, respectively. 
$\hat n^{(\beta)}_l=\hat\beta^{\dagger}_l\hat\beta_l$ is the number operator, with band index $\beta \in \{a,b\}$ for the lower and the upper band, respectively.

\item[$(iii)$] The parameter space, defined by the parameters $\{J_{\beta},W_{\beta},W_x,C_{\mu},\Delta\}$, is quite large.  
$J_{\beta}$ is the kinetic energy scale, characterizing the nearest neighbor hopping matrix elements. The $W$'s represent the on-site, intra- and inter-band interaction strengths. 
$2\mu+1$ dipole coupling coefficients $C_{\mu}$ induced by the constant field $\omega_B\equiv 2\pi F$, with $\mu\in \mathbb{Z}$. Hence, the Stark force (the tilt) is given by the third term
in the first line of Eq.~(\ref{eq:01}). The average band distance, the energy band gap, is given by $\Delta$. 
All these parameters are expressed in the following in units of the recoil energy $E_{\rm rec}=(\hbar k_L)^2/2m$, 
which a photon with momentum $\hbar k_L$ exchanges when scattering from an atom of mass $m$. Since all potentials 
are created optically in the experiment, this is the natural scale for realizations with ultracold atoms \cite{Ibloch,Mors06}. 
Furthermore, we set $\hbar=1$ throughout this paper.

\item[$(iv)$] The available state space $\mathcal F$ is spanned by the following Fock states
\begin{eqnarray}\label{eq:03}
\mathcal F = {\rm  span}\{|n_1^an_2^a\cdots\rangle\otimes|n_1^bn_2^b,\cdots\rangle,\cdots\},
\end{eqnarray}
with $\sum_ln^a_l+n^b_l=N$. This space can be split into sets of states with the same upper-band occupation number $M=\langle \sum_ln^b_l\rangle_{\phi}$, with $|\phi\rangle$ an arbitrary Fock state. We refer to these sets of states as $M$-Manifolds in the following. In this way, after relabeling the states using the quantum number $M$, the Hamiltonian matrix takes the following block form
\begin{equation}\label{eq:04}
\mathbf H=\left[
\begin{array}{cccccc}
 \mathbf{H}_{0,0} &  \mathbf{H}_{1,0} & \mathbf{H}_{2,0}   & \mathbf 0&\cdots & \mathbf 0\\
 \mathbf{H}_{0,1} &  \mathbf{H}_{1,1} &                   & \ddots & &\vdots \\
 \mathbf{H}_{0,2} &                  & \ddots            & \ddots &\ddots & \mathbf 0\\
  \mathbf{0}     &  \ddots          & \ddots            & \ddots&& \mathbf{H}_{N,N-2}\\
  \vdots         &                  &  \ddots           & & \mathbf{H}_{N-1,N-1} & \mathbf{H}_{N,N-1}\\ 
  \mathbf 0      &  \cdots          & \mathbf 0         & \mathbf{H}_{N-2,N}&\mathbf{H}_{N-1,N} & \mathbf{H}_{N,N} \\ 
\end{array}
\right]\,.
\end{equation}
The off-diagonal blocks represent the coupling of the different $M$-subsets via one- and two-particle exchange processes induced by the Hamiltonian terms in the second line of Eq.~(\ref{eq:01}).

In the non-interacting limit, i.e. $W_{\beta,x}=0$, the matrix $\mathbf H$ can be effectively reduced to a block-tridiagonal matrix of size $(N+1)\times (N+1)$ with approximate
eigenvalues $E_{M}\approx M\Delta_r$, see ref.~\cite{Carlitos02,PhDThesis2013}. Here 
\begin{eqnarray}\label{eq:05}
 \Delta_r=\Delta\sqrt{\left(1-\omega_B r/\Delta\right)^2+4(\omega_B C_0/\Delta)^2}
\end{eqnarray}
is the energy difference between neighboring manifolds exchanging one particle. The integer $r$ labels the
order of the resonances \cite{Carlitos01} occurring for tilts $F_r\approx \Delta /2\pi r$. These resonances appear at certain values of the force for which the single-particle levels in the
upper band become degenerate with the ones of the lower band \cite{Korsch,Ploetz}. The region in parameter space as a function of $F$ close to $F_r$ is denoted as resonant tunneling (RET) regime in the following. The interactions, together with the strong interband coupling at RET, lead to a strong mixing of levels. In contrast, we can resort to a perturbative argument far off the resonances. Off-resonance,
the eigenvalues $E_M$ are corrected by including the splitting of the single particle levels due to the interactions. The corresponding eigenstates can be labeled by the integers $\theta_{\beta=1,2}=\langle\frac{1}{2}\sum_ln^{\beta}(n^{\beta}-1)\rangle_{\phi}$ and $\theta_x=2\langle\sum_ln^a_ln^b_l\rangle_{\phi}$. Therefore, they can be approached by the formula \cite{Carlitos01,Carlitos02}:
\begin{eqnarray}\label{eq:06}
 E_{M,\theta} =  M\Delta_r + \vect{\theta}\cdot\mathbf{W}, 
\end{eqnarray}
with $\mathbf{W}^T=(W_a,W_b,W_x)$, and $\vect{\theta}=(\theta_a,\theta_b,\theta_x)$. The maximal intramanifold splitting generated by the interaction can be estimated by $\delta\varepsilon={\rm max}\{U^M_{a,b,x}\}$, with
\begin{eqnarray}\label{eq:07}
  (U^M_a)_{\rm max}&=&\frac{W_a}{2}(N-M)(N-M-1),\nonumber\\
  (U^M_b)_{\rm max}&=&\frac{W_b}{2}M(M-1),\nonumber\\
  (U^M_{ab})_{\rm max}&=&2W_x(N-M)M\,.
\end{eqnarray}
\begin{figure}[t]
\centering 
\includegraphics[width=0.65\columnwidth]{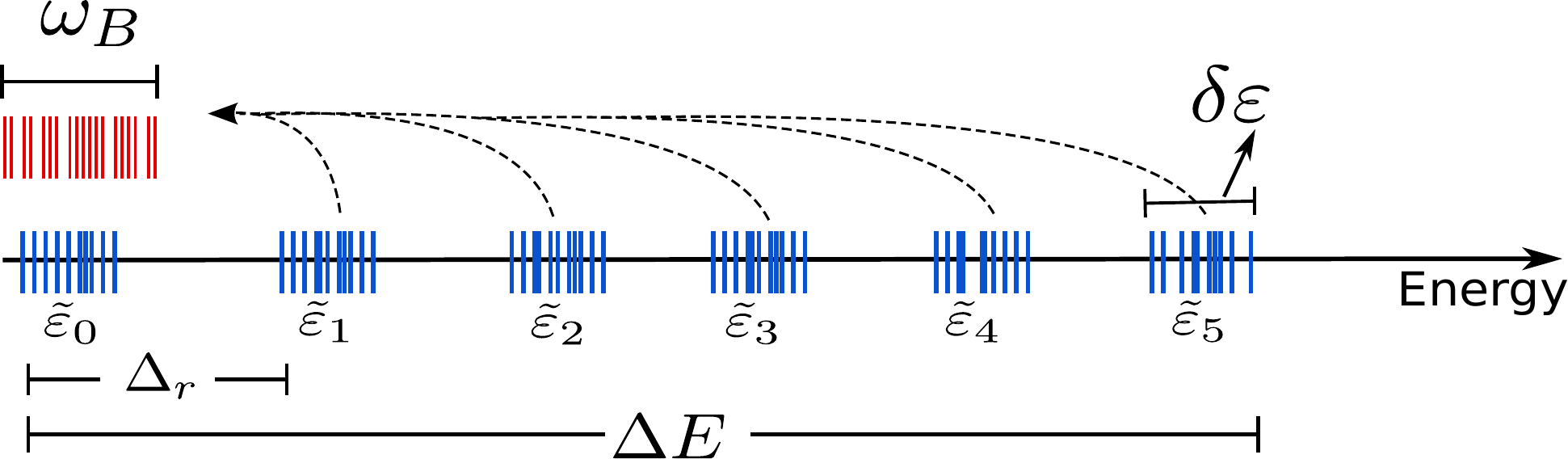}
\caption{\label{fig:00}(Color online) Sketch of the structure of eigenvalues for a fixed Stark field $F$ far from the resonance regime, 
  for the tilted Hamiltonian in Eq.~(\ref{eq:01}) (lower/blue bars), and of the Floquet quasienergies, i.e. the folded spectrum within a Floquet zone of width $\omega_B$ 
  (upper/red bars), see sections \ref{sec:02} and ~\ref{sec:03-1} for details. The values $\tilde{\varepsilon}_i$ are the Lanczos starting points used for improving the 
  performance of the diagonalization as presented in section \ref{sec:03-1}. $\delta\varepsilon$ is the maximal splitting of single-particle states generated by the interparticle interaction.
}
\end{figure}

The values of $U^M_{a,b,x}$ correspond to the largest energy cost for producing states with $M$ particles in a single lattice site, e.g., 
$|\phi\rangle\sim|N-M00,\cdots\rangle\otimes|M00,\cdots\rangle$. In consequence, in most off-resonant cases with $F$ far from $F_r$, we have $\Delta_r\sim\Delta\gg W_{a,b,x}$, and, therefore, the overlapping between contiguous $M$-manifolds can be neglected. Then, the spectrum consists of bunches of levels within an energy range $\delta\varepsilon$ around $E_M$, as sketched in Fig.~\ref{fig:00}. 

According to Eq.~(\ref{eq:06}), the intramanifold energies $E_{M,\theta}$, far from the $F_r$, are proportional to the tilt, c.f. also Fig.~\ref{fig:00-0}(a). Therefore, the parametric dependence of the levels on $F$, $\varepsilon(F)$, is essentially given by straight lines parallel to each other, with slopes $\approx -M\Delta/F_r$. Conversely, the intermanifold energy levels approach each other in the vicinity of the interband resonances at $F=F_r$ (see Fig.~\ref{fig:00-0}). Here avoided crossings take place involving all the levels arising from the lack of additional symmetries and from the strong interband couplings \cite{Carlitos01}. Within the resonant regime, Eq.~(\ref{eq:06}) is clearly no longer valid. The eigenenergies lose their manifold structure since typically we have $\Delta\gg W_{a,b,x}\sim \Delta_r$. Nevertheless, Eq.~(\ref{eq:06}) is still useful for performance tests of the numerical routines presented in the next sections.

\item[$(v)$] 
A helpful trick for the numerical solution of the Hamiltonian (\ref{eq:01}) is to transform it into the interaction picture
with respect to the tilting term $\sum_{l,\beta}\omega_Bl \hat n_l^{(\beta)}$. This procedure allows us to recover
the translational invariance in the new Hamiltonian $H'$. We can, therefore, impose periodic boundary conditions in space, i.e., 
$\hat{\beta}^{\dagger}_{L+1}=\hat{\beta}^{\dagger}_{1}$. Thereby, we now work with the common eigenbasis of the translation 
operator $\hat S$ and the transformed Hamiltonian $\hat H'$ since $[\hat H',\hat S]=0$. 
This new basis is the translationally invariant Fock (TIF) basis indexed by $\alpha$ and introduced, e.g., in ref.~\cite{kol68PRE2003}:
\begin{eqnarray}\label{eq:08}
|s_{\alpha},\kappa\rangle=D_{\alpha}^{-1/2}\sum^{D_{\alpha}}_{l=1}e^{i2\pi\kappa l}\hat S^l |n_1^an_2^a\cdots\rangle\otimes|n_1^bn_2^b,\cdots\rangle_{\alpha}\,,
\end{eqnarray}
\noindent where $\hat S^m|s_{\alpha},\kappa\rangle=e^{-i\omega_B  \kappa m}|s_{\alpha},\kappa\rangle$. The action of the
translational operator on a Fock state is $$\hat S^m|n_1\cdots n_l\cdots\rangle=|n_{1+m}\cdots n_{l+m}\cdots\rangle\,.$$
$\kappa \equiv \kappa_j=j/D_{\alpha}$ ($j\in[0,D_{\alpha}-1]$) is the lattice quasimomentum and $D_{\alpha}$ is the total number of
cyclical permutations of the Fock state. The dimension of the Hilbert space expanded by the TIF basis is $d_s\approx d_F/L$.
It is easily seen that the block structure of the Hamiltonian matrix remains invariant and the $M$-sets, now $\{|s_{\alpha},\kappa,M\rangle\}$,
have dimensions
\begin{eqnarray}\label{eq:09}
d_M=\frac{1}{L}{M+L-1\choose L-1}{N-M+L-1\choose  L-1}\,,
\end{eqnarray}
where $(\cdots)$ stands for the combinatorial function. The full dimension of the Hilbert space is then given by $d_s=\sum_Md_M$. 

\begin{figure}[t]
\centering 
\includegraphics[width=0.7\columnwidth]{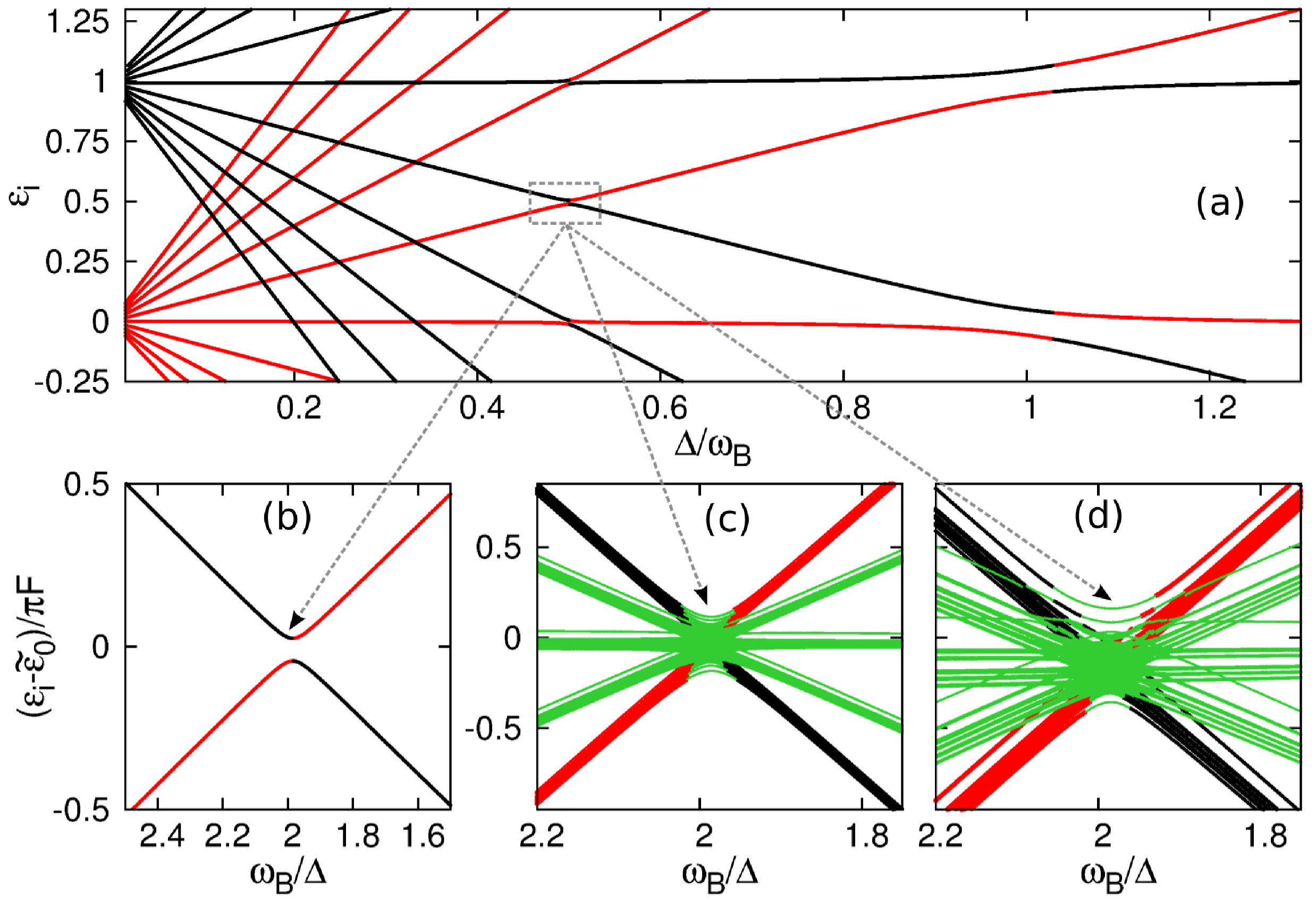}
\caption{\label{fig:00-0}(Color online) Wannier-Stark spectrum for the single-particle case (a,b) and typical system parameters. 
Panels (c) and (d) show the resonant regime for $N/L=4/5$ and weak ($W_a=W_b=1.3\times10^{-3}, W_x=1.4\times10^{-3}$) or strong
($W_a=W_b=1.3\times10^{-2}, W_x=1.4\times10^{-2}$) interparticle interactions, around the second order resonance at $\omega_B/\Delta =2$
for $J_a=0.059, J_b=-0.071, C_0=-0.095, C_1=0.043, C_2=-0.005, \Delta=0.56$.
The color code is as follows: eigenstates with $M_i\simeq 0$ (black lines), eigenstates with $M_i\simeq N$ (red/dark grey lines), and eigenstates with $0<M_i<N$ (green/light grey lines). 
For better visibility of the avoided crossings, we plot the panels (b-d) directly as a function of the Stark force $F \propto \omega_B$, 
whilst the fan of single-particle Wannier-Stark levels in (a) is best seen as a function of the inverse Stark force $\propto 1/\omega_B$.
}
\end{figure}

A second important consequence of transforming $\hat H$ into $\hat H'$ is that the new Hamiltonian is explicitly time dependent. 
This originates in the transformed terms  
\begin{eqnarray}\label{eq:10}
\hat{\beta}^{\dagger}_{l+1}\hat{\beta}_l&\rightarrow&\hat{\beta}^{\dagger}_{l+1}\hat{\beta}_l\exp{(-i\omega_Bt)}\nonumber\\ 
\hat{a}^{\dagger}_{l+\mu}\hat{b}_l&\rightarrow&\hat{a}^{\dagger}_{l+\mu}
\hat{b}_l\exp{(-i\omega_B \mu t)} \,. 
\end{eqnarray}
One can easily see that $\hat H'(t)$ is periodic in time with a fundamental period $T_B=2\pi/\omega_B$, the Bloch period. The temporal periodicity
allows one to obtain a stationary spectrum, not of the Hamiltonian, but of the single-cycle Floquet operator \cite{Shirley1965,kol68PRE2003}
\begin{eqnarray}\label{eq:11}
\hat U(T_B)=\hat \mathcal T \exp\left(-i \int_0^{T_B} dt \, \hat H'(t)\right) \,,
\end{eqnarray}
where $\hat\mathcal T$ implies time ordering. The eigenphases $\phi_j$ 
\begin{eqnarray}\label{eq:12}
\hat U(T_B)|\phi_j\rangle=e^{-i\phi_j}|\phi_j\rangle\,,
\end{eqnarray}
divided by the period $T_B$ define the quasienergies, in analogy to quasimomentum for the Bloch
problem of a particle in a periodic potential. The set of quasienergies $\{\varepsilon_j\}$ inherits the periodicity $\omega_B$ in energy space. This is
a direct consequence of the Floquet theorem for periodically time-dependent systems \cite{Shirley1965,Buch1993,BuchJosaB1995}.
Therefore, we can choose, without loss of generality, $\varepsilon_j\in(-\omega_B/2,\omega_B/2]$, implying that the spectrum is
folded within an energy range of size $\omega_B$, the fundamental Floquet zone (FZ), as sketched by the red bars in Fig.~\ref{fig:00}.

The so-called extended spectrum is obtained by shifting the eigenenergies from FZ as $\varepsilon_j\rightarrow\varepsilon_j+n_{f}\omega_B$, 
with the integer $n_f$ labeling the FZ. To obtain numerically the spectrum directly from $\hat U(T_B)$ typically demands long
computation times since the often applied approach consists of two steps. First, one computes the matrix $\hat U(T_B)$ by propagating the
entire basis, for instance, the TIF basis, from $t=0$ to $t=T_B$. Subsequently, the obtained matrix must still be diagonalized, whereby
it is not a sparse but a full matrix in general \cite{kol68PRE2003,Tomadin}.  In this paper, we follow a different but equivalent approach: 
We study and diagonalize the Floquet Hamiltonian to be introduced in the next section. 
This procedure avoids the time integration for obtaining the one-cycle operator from Eq.~(\ref{eq:11}), as done in previous single-band studies of
many-body Wannier-Stark problems \cite{kolo03,Tomadin,kol68PRE2003}.
Now we directly diagonalize much larger, but sparse matrices instead. We shall use the properties of our Hamiltonian (\ref{eq:01}) 
described above in order to improve the performance of the diagonalization scheme that will be presented in the next section.

\end{itemize}

\section{Floquet-Lanczos Diagonalization: Generalities and performance}
\label{sec:03}

In this section we discuss the implementation of two well-known techniques, Floquet theory and Lanczos diagonalization. The combination of these two methods allows for a detailed numerical treatment of the Hamiltonian $\hat H'$ describing our many-body Wannier-Stark problem for reasonable system sizes.

\subsection{Floquet theory}
\label{sec:03-1}

Let us start with the Floquet representation of transformed $\hat H'(t)$. The periodicity in time of $\hat H'(t)$, i.e.,  
$\hat H'(t+T_B)=\hat H'(t)$, permits the application of the Floquet theorem \cite{Shirley1965,Buch1993,BuchJosaB1995}.
The Floquet Hamiltonian, defined as $\hat H_f=\hat H'(t)-i\partial_t$, has instantaneous eigenvectors
\begin{equation}\label{eq:13}
\hat H_f|\varepsilon_j(t)\rangle=\varepsilon_j|\varepsilon_j(t)\rangle \,,
\end{equation}
which have the same periodicity as $\hat H'(t)$, i.e. $|\varepsilon_i(t+T_B)\rangle=|\varepsilon_i(t)\rangle$. 
Any time-dependent solution can be expressed now as
\begin{equation}\label{eq:14}
|\psi_t\rangle = \sum_j c^0_j \exp\left(-i\varepsilon_j t\right)|\varepsilon_j(t)\rangle \,,
\end{equation}
where the coefficients $c^0_j$ are the projections of the initial state onto the Floquet basis vectors.

In order to obtain a system of stationary equations, which we can solve by diagonalization, we expand these vectors into their Fourier series (making perfect sense because of their
periodicity). The corresponding multi-mode Fourier decomposition reads
\begin{equation}\label{eq:15}
|{\varepsilon_j}(t)\rangle=\sum_{\vec{k}} e^{-i\vec{k}\cdot\vec{\omega}t} 
|\phi^{\vec{k}}_{\varepsilon_j}\rangle= \sum_{k} e^{-ik\omega_Bt}|\phi^{k}_{\varepsilon_j}\rangle \,.
\end{equation}
The expression on the very right is possible because of the existence of a fundamental frequency $\omega_B$ since also the remaining 
frequencies are multiples of $\omega_B$; so we can write: $\vec k\cdot\vec\omega=(k_1+2k_2+...+\mu k_{\mu})\omega_B=k\omega_B$. 
The eigensystem of equations is then given by the following coupled algebraic equations:
\begin{eqnarray}\label{eq:16}
\varepsilon_j\hat{1}|\phi^{k}_{\varepsilon_j}\rangle&=&
\left({\hat H_0}- \omega_B k\hat{1}\right)|\phi^{k}_{\varepsilon_i}\rangle+{\hat J}\arrowvert \phi^{k-1}_{\varepsilon_j}\rangle
+{\hat J}^{\dagger}\arrowvert \phi^{k+1}_{\varepsilon_j}\rangle\nonumber\\
&+&\sum_{\mu}\left[{\hat Z}_{\mu}\arrowvert \phi^{k-\mu}_{\varepsilon_j}\rangle
+{\hat Z}_{\mu}^{\dagger}| \phi^{k+\mu}_{\varepsilon_j}\rangle\right]\,.
\end{eqnarray}
Here, $\hat H_0$ collects all the time-independent parts of $\hat H'(t)$. The operators $\hat{J}$ and $\hat Z_{\mu}$ define the hopping and 
dipole-like transition operators
\begin{eqnarray}\label{eq:17}
\hat J&=&-\frac{1}{2}\sum_{l,\beta}J_{\beta}\hat{\beta}^{\dagger}_{l+1}\hat{\beta}_{l}\nonumber\\
\hat Z_{\mu}&=&\omega_B C_{\mu}\sum_l\hat{a}^\dagger_{l+\mu}\hat{b}_l\,.
\end{eqnarray}
Equation (\ref{eq:16}) can be rewritten in the matrix form  
\begin{eqnarray}\label{eq:18}
(\mathcal{M}-\varepsilon_j \mathcal{B})\cdot\mathbf{x}_j=0\,,
\end{eqnarray}
after expanding the states of the $k$th Fourier component $|\phi^{k}_{\varepsilon_i}\rangle$ in the TIF basis,   
$|\phi^k_{\varepsilon_j}\rangle= \sum_{\alpha} C^j_{k\alpha}|s_{\alpha}\rangle$, and defining the vector of components 
$\mathbf{x}^T_j=\left(C^j_{k_{\rm  max},\alpha},\cdots,C^j_{k,\alpha},\cdots,C^j_{k_{\rm min},\alpha}\right)$. In our case, $\mathcal{B}=\mathbf{1}$, 
giving a standard eigenvalue problem. The resulting square matrix is banded and blocked and it has the total dimension 
\begin{equation}\label{eq:19}
n_{\rm tot} = d_s \Delta k \,,
\end{equation}
where $\Delta k \equiv k_{\rm max}+k_{\rm min}+1$ is the total number of Fourier components considered, with $-k_{\rm min}\leq k \leq k_{\rm max}$. 
The matrix $\mathcal M$ is thus given by
\begin{equation}\label{eq:20}
\mathcal{M}=
\left[
\begin{array}{ccccccccc}
\mathbf{H}_0+k_{\rm max}\omega_B\mathbf{1}&&&&&&&&\\
&\ddots&\vdots&&&&&&\\
&\cdots&\mathbf{H}_0+2\omega_B\mathbf{1} & \mathbf{J}^{\dagger}+\mathbf{Z}^{\dagger}_1& \mathbf{Z}_2^{\dagger}&\mathbf{Z}_3^{\dagger} &\iddots&&\\
&&\mathbf{J}+\mathbf{Z}_1&\mathbf{H}_0+\omega_B\mathbf{1} & \mathbf{J}^{\dagger}+\mathbf{Z}^{\dagger}_1& \mathbf{Z}_2^{\dagger}&\mathbf{Z}_3^{\dagger}&&\\
&&\mathbf{Z}_2&\mathbf{J}+\mathbf{Z}_1& \mathbf{H}_0 & \mathbf{J}^{\dagger}+\mathbf{Z}^{\dagger}_1&\mathbf{Z}_2^{\dagger}&&\\
&&\mathbf{Z}_3&\mathbf{Z}_2 & \mathbf{J}+\mathbf{Z}_1&\mathbf{H}_0-\omega_B\mathbf{1}& \mathbf{J}^{\dagger}+\mathbf{Z}^{\dagger}_1&&\\
&&\iddots&\mathbf{Z}_3&\mathbf{Z}_2 &\mathbf{J}+\mathbf{Z}_1&\mathbf{H}_0-2\omega_B\mathbf{1}&\cdots&\\
&&&&&&\vdots&\ddots&\\
&&&&&&&&\mathbf{H}_0-k_{\rm min}\omega_B\mathbf{1}\\
\end{array}
\right].
\end{equation}
The matrix $\mathcal M$ inherits the sparseness of $\mathbf H$; the number of non-diagonal blocks depends on the number of considered $\mu$-dipole couplings. In the current case, we can restrict to $|\mu|\leq 2$, for which $\mu=\{-2,-1,0,1,2\}$ since the dipole coefficients go fast to zero with increasing $|\mu|$ \cite{PhDThesis2013}. As a consequence, the matrix $\mathcal M$ is effectively a very sparse, five-diagonal block matrix. As for the operator $\hat U(T_B)$, the eigenvalues of the Floquet matrix lie within 
the fundamental Floquet zone $\varepsilon_j\in (-\omega_B/2,\omega_B/2]$, with $\varepsilon_j{\rm mod}(\omega_B)$. The eigenvectors are 
obtained as the linear combination $|\varepsilon_j\rangle=\sum_{k\alpha}C^j_{k\alpha}|s_{\alpha}\rangle=\sum_{\alpha}C^j_{\alpha}|s_{\alpha}\rangle$,
with $\sum_{k}C^j_{k\alpha}=C^j_{\alpha}$. The dimension of the energy basis of interest, $\{|\varepsilon_j\rangle\}_j$, is therefore $d_s$.

Since a maximal number $d_s\,(\ll n_{\rm tot})$ of eigenenergies lie within one FZ, the computation of the full set of eigenvalues of $\mathcal M$ ($n_{\rm tot}$) is not necessary. 
Therefore, it is convenient to implement a sophisticated algorithm that permits us to compute a controllable number of eigenvalues and eigenstates of $\mathcal M$. 
We implemented the symmetric Lanczos algorithm, which proved to be very efficient for obtaining either all or at least a reasonable number of eigenvalues in the fundamental FZ. 
Without loss of generality, we work with the TIF subspace defined by quasimomentum $\kappa=0$ (see Sec.~\ref{sec:02}). This choice automatically guarantees $\mathcal M$ to be a real and symmetric matrix.

\subsection{Lanczos diagonalization}
\label{sec:03-2}

The sparseness of the matrix $\mathcal M$ makes it very suitable for numerical diagonalization by means of algorithms such as the symmetric Lanczos procedure 
\cite{GeneBook}. This method allows us the computation of eigenvalues and eigenstates of a given complex symmetric matrix. A nice feature of this method is that it
permits the computation of eigenvalues close to a given initial choice in energy space. 
Let us start by shifting the diagonal elements as $\mathcal M\rightarrow\mathcal M+\tilde{\varepsilon}_0\mathbf{1}$. The goal is to find the 
largest eigenvalues $\lambda_j=1/(\varepsilon_j-\tilde{\varepsilon}_0)$ of the matrix $\mathcal{M}^{-1}$, 
which correspond to the eigenvalues $\varepsilon_j$ close to the predefined shift parameter $\tilde{\varepsilon}_0$. 

The method is an adaptation of the power method \cite{GeneBook} to find eigenvalues and eigenvectors of a square matrix  
based on the construction of the Krylov subspace $(\mathcal K_{n_p})$; that is, the space spanned by
\begin{eqnarray}\label{eq:21}
 \mathcal K_{n_p}[\mathcal{M}^{-1},\eta_1]={\rm span}\{\mathbf{\eta}_1,\mathbf{\eta}_2,\cdots,\mathbf{\eta}_{n_p}\},
\end{eqnarray} 
with $\mathbf{\eta}_{j+1}\equiv(\mathcal{M}^{-1})^{j}\eta_1$. $\eta_1$ is a suitable initial vector, which is normalized 
$\|\eta_1\|=1$. $n_p\;(\ll n_{\rm tot})$ is the dimension of the Krylov subspace, here
referred to as Lanczos iterations number. We define the matrix $\mathcal{P}=\left[\eta_1,\eta_2,...,\eta_{n_p-1}\right]$. 
The Krylov subspace representation of  $\mathcal{M}^{-1}$ is the matrix $\mathcal{T}=\mathcal{P}^{-1}\mathcal{M}^{-1}\mathcal{P}$, 
which is tridiagonal and symmetric, with elements given by 
\begin{eqnarray}\label{eq:22}
\alpha_j&=&\mathcal T_{j,j}=\eta^T_j\mathcal M^{-1} \eta_j,\nonumber\\
\beta_j&=&\mathcal T_{j,j+1}=\|\mathbf t^{T}_{j+1} \mathbf t_{j+1}\|\,,
\end{eqnarray}
with 
\begin{eqnarray}\label{eq:23}
\mathbf t^{T}_{j+1}&=&\mathcal M^{-1} \eta_j-\alpha_j\eta_j-\beta_{j-1}\eta_{j-1}\nonumber\\
\eta_{j+1}&=&\mathbf t_{j+1}/\beta_{j+1}.
\end{eqnarray}
The construction of the $\mathcal K_{n_p}$ space is supported by an appropriate Gram-Schmidt orthogonalization procedure for the $\eta_i$ vectors.
The matrix $\mathcal T$ can be diagonalized using standard diagonalization routines. Here we use the QR algorithm \cite{GeneBook} with an error of the order of the numerical (double) precision $10^{-14}$; that is, we compute the decomposition $\mathcal T = QR$ and the subsequent iterations 
\begin{eqnarray}\label{eq:24}
\mathcal T_{j+1}=R_jQ_j=Q^T_j\mathcal T_jQ_j=Q^{-1}_j\mathcal T_jQ_j\,,
\end{eqnarray}
with $\mathcal T_0=\mathcal T$. The matrix $\mathcal T_{j+1}$, by the equation above similar to $\mathcal T$, converges to the Schur
form of symmetric matrix $\mathcal T$, i.e., the eigendecomposition of $\mathcal T$. In addition,
the discrimination criterion introduced by Parlett and Scott \cite{ParletScott1979} is implemented in order to separate
converged from non-converged eigenvalues. Finally, since the matrices $\mathcal T$ and $\mathcal M^{-1}$ are similar, they have the same eigensystem.

The numerical implementation of the Lanczos algorithm requires the setting of several parameters. The first of these is the shift $\tilde{\varepsilon}_0$,
which we set using Eq.~(\ref{eq:06}), for instance, $\tilde{\varepsilon}_0=E_M$. This allows us to find a set of
eigenvalues in the vicinity of manifold energy $E_M=M\Delta_r$ of interest to us (see Fig.~\ref{fig:00}). Using the approximately known dependence of 
$\tilde{\varepsilon}_0 (F)$ on the Stark force $F$ permits us to set the center of the Floquet zone, for instance, in middle of the spectral range, 
$\tilde{\varepsilon}_0\approx\frac{1}{2}N\Delta$, for convenience. This choice helps in the visualization of the spectrum in the vicinity of $F_r$, as shown for instance in Fig.~\ref{fig:00-0}.

The number of the found eigenvalues depends on the number of  Fourier components $\Delta k$ and the number of 
Lanczos iterations $n_p$. $k_{\rm max}$ can be estimated by the ratio of the eigenenergy bandwidth  
of $\mathbf H$ in Eq.~(\ref{eq:04})
and the size of the FZ $\omega_B$.  
In this way, $k_{\rm max}$ can be regarded as the number of Stark-induced ''fictitious" photonic excitations $\hbar \omega_B$ necessary to promote all atoms from the lowest 
($M=0$) to the largest ($M=N$) manifold excitation (see the sketch in Fig.~\ref{fig:00}). This value is thus given by
\begin{equation}\label{eq:25}
 k_{\rm max}=\frac{\Delta E}{\omega_B} \approx \frac{N\Delta_r+W_bN^2}{2\pi F},
\end{equation}
where $W_bN^2$ is the energetic cost of producing a state with $N$ atoms in a single lattice site within the second band \cite{Carlitos01}, e.g., the state 
$|\varepsilon_j\rangle\sim |000\cdots\rangle\otimes|N00\cdots\rangle$. For our purpose, it is very important to note that according to the estimation of 
$k_{\rm max}$, we have that $k_{\rm min}$ does just play a minor role. Therefore, the Fourier expansion may not necessarily be
symmetric with respect to the photon index $k$, i.e. $k_{\rm min}/k_{\rm max}<1$ is indeed a possible choice for us. To minimize the dimension of $\mathcal M$, we may even choose $k_{\rm min}=1$ and still obtain converged Floquet eigenspectra. The number of Lanczos iterations $n_p$ is adapted such that we obtain a minimum number $d_s$ of convergent eigenvalues within the FZ, for which we usually have that $d_s<n_p\ll n_{\rm tot}$. 

\begin{figure}[t]
\centering 
\includegraphics[width=0.7\columnwidth]{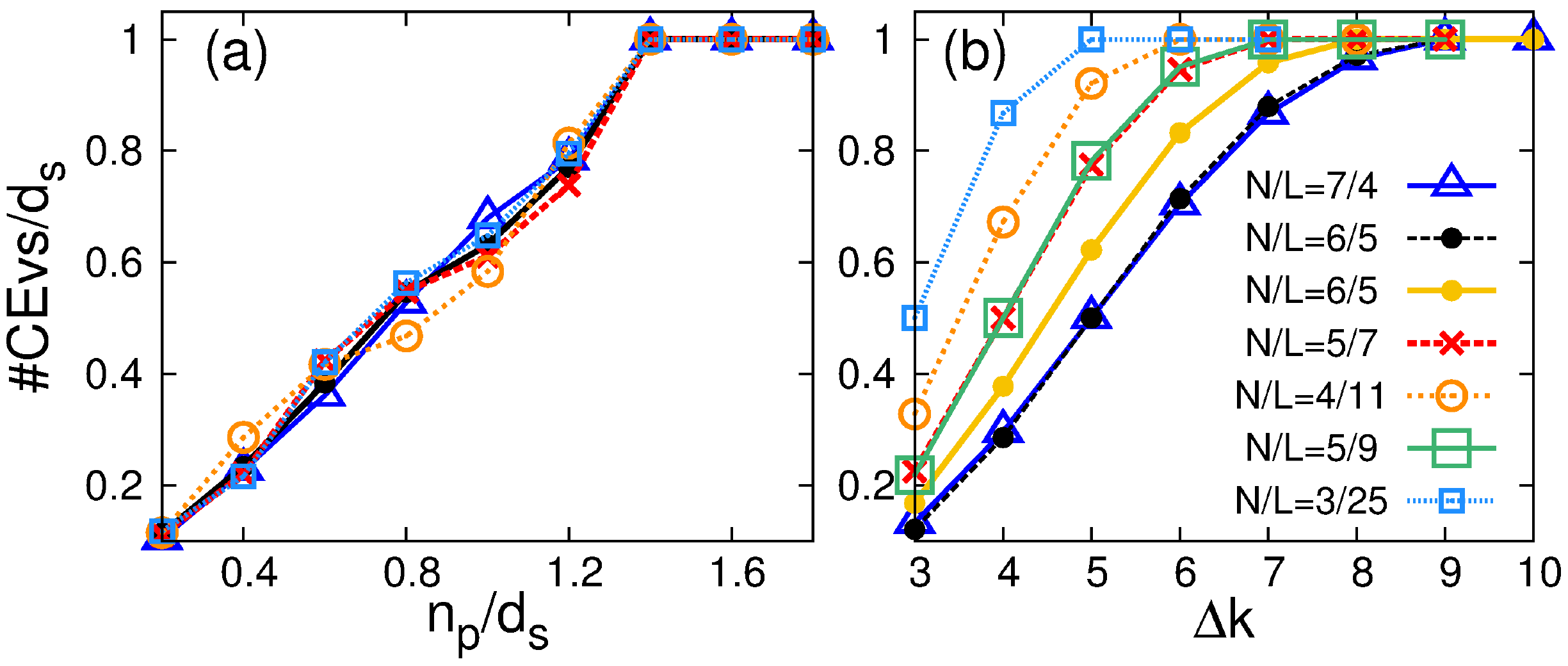}
\caption{\label{fig:01-0}(Color online) 
Spectra computed at the first order resonance between the two energy bands for increasing sizes of our system. i.e. for the filling factors shown in the legend in (b).
Panel (a) shows the saturation of the number of convergent eigenenergies ($\#$CEvs), within one Floquet zone (FZ), with increasing size of the Krylov subspace. The $y$-axis is normalized to the corresponding TIF basis dimension, $d_s$ and $k_{\rm max}$, are estimated as in  Eq.~(\ref{eq:25}). Panel (b) shows the saturation of $\#$CEvs with the number of Fourier components $\Delta k$. The realistic parameters \cite{Carlitos01,PhDThesis2013} for this computation are: $r=1$, $\Delta=2.53$, $J_a=0.08$, $J_b=-0.24$, $W_a=0.018$, $W_b=0.025$, $W_x=0.02$, $C_0=-0.09$, $C_{\pm 1}=0.035$, and $C_{\pm 2}=0.002$.}
\end{figure}

\begin{figure}[t]
\centering 
\includegraphics[width=0.7\columnwidth]{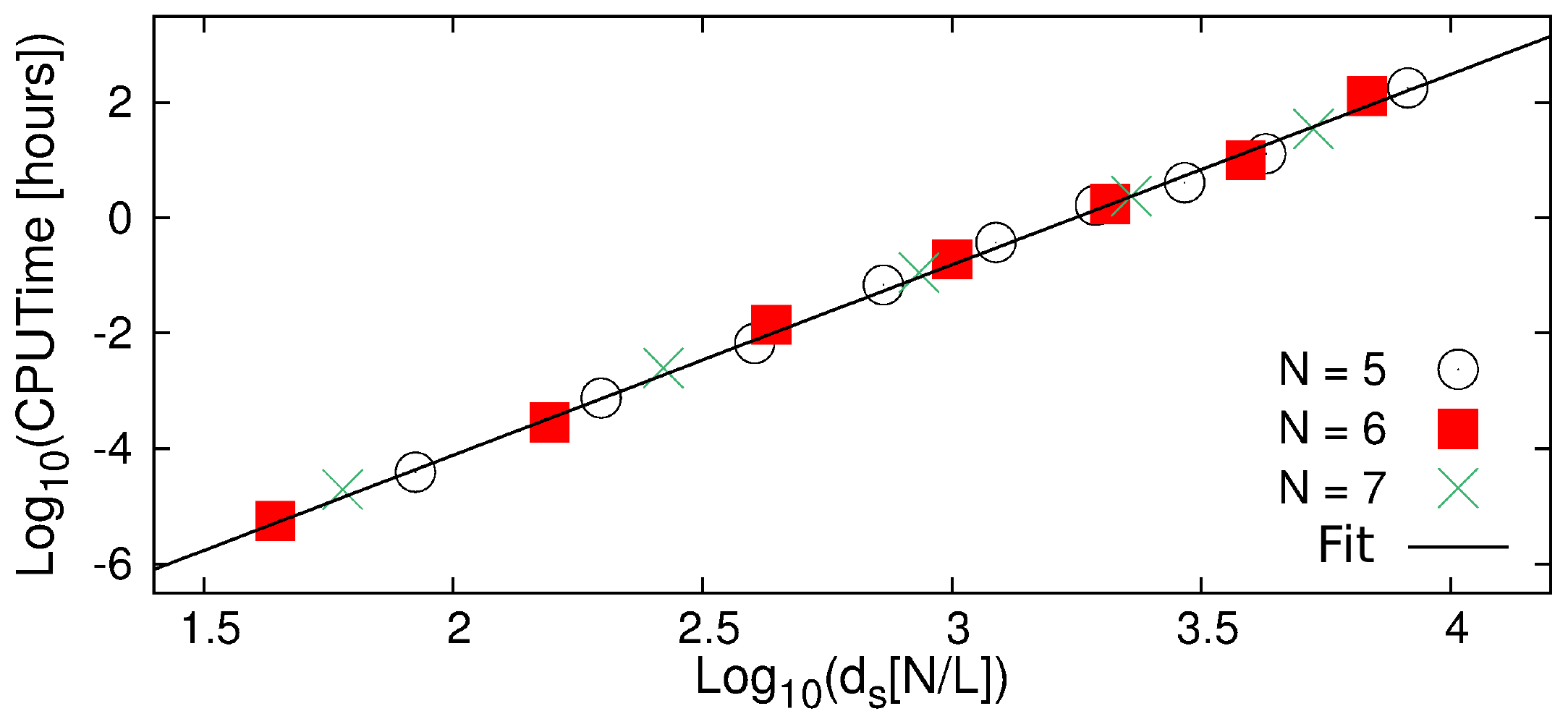}
\caption{\label{fig:01-1}(Color online) Log-log plot of the computational time (CPU time) vs. the TIF basis
 dimension for the quasimomentum block with $\kappa=0$. We extract the following relation for the CPU time $\sim d_s^{n}$
 with $n=3.2\pm 0.2$. The exponent $n$ is estimated by the slope of the fitting function for the data corresponding to $N=5$ (black $\circ$), given by the black straight line. 
 Note that for $N=\{6,7\}$ no relevant changes occur as one can see even by eye. $d_s$ is increased by fixing the total particle number $N$ and varying the number of lattice sites over one order of magnitude $L=\{3,\cdots,12\}$. The dimensions of the corresponding Floquet matrices $\mathcal M$ are then $n_{\rm tot}\sim 10^2 \cdots 10^5$. The system parameters are the same as applied in Fig.~\ref{fig:01-0}.
}
\end{figure}

The required memory storage for one diagonalization is given by the formula 
\begin{equation}\label{eq:26}
 {\rm MS[GB]}=\left(n_{\rm tot}n_{\rm larg}+n_p^2\right)\times\frac{16}{10^9}\sim d_s^2 \,,
\end{equation}
where $n_{\rm larg}=(\mu+1)d_s$ and $n_p\equiv{\rm dim}(\mathcal K_{n_p})$.  The first term on the right hand side is the needed memory 
storage for $\mathcal M$, which for systems with fillings, i.e. particles per number of sites, $N/L>1$ can be very large. $n_p$ is the dimension of the Krylov matrix $\mathcal P$,
 which is proportional to the dimension of the TIF basis. For storing a 64bit number we need 8 bytes. The factor $2 \times 8=16$ comes from the fact that we use an implementation of the Lanczos algorithm which was developed for general complex symmetric matrices \cite{Buch1993,BuchJosaB1995,Krug}. For instance, for $N/L=7/6$ ($d_s=5304$) we need $k_{\rm max}> N/r$ Fourier components at RET conditions $F=F_r$, given $\Delta\gg \{W_{\beta,x},C_{\mu}\}$. Therefore, we have in this case that ${\rm MS[GB]}\sim 35$. The computer used for these calculations was a 24 core AMD Opteron(TM) Processor 6234 with 60-GB main memory.

In Fig. \ref{fig:01-0}-(a), we show the dependence of the number of convergent eigenenergies of $\mathcal M$, normalized to $d_s$, 
on the size of the Krylov subspace, for different filling factors $N/L$. Once $k_{\rm max}$ is set by Eq.~(\ref{eq:25}), we see that
the saturation point $(n_p/d_s\approx 1.3)$ is the same for all systems. In Fig.~\ref{fig:01-0}-(b), we scan instead the number of Fourier components 
by varying $k_{\rm max}$ for fixed size of the $\mathcal K_{n_p}$ subspace $n_p=2 d_s$. Figure~\ref{fig:01-1} shows a polynomial growth of
the computational time (CPU time) with increasing dimension of the TIF basis. We observe a polynomial increase of the CPU consistent with $\sim d_s^3$. 
Indeed the only limitation in our numerical scheme is the memory storage since  $k_{\rm max}\sim \omega^{-1}_B = 1/2\pi F$, which implies that 
the dimension of the $\mathcal M$ increases considerably for smaller and smaller values of $F$. This fact makes it difficult to study, for instance, high order resonances with large $r$, for which the tilt is small. A positive aspect of our system is, however, that many of its properties are not really size-dependent, but they rather depend only on the ratio $N/L$ (see ref.~\cite{Carlitos01}). The growth law for the computational time and, in consequence, the efficiency of our numerical method are expected to hold for even large systems as may be extrapolated from Fig.~\ref{fig:01-1}.  

One way to overcome the memory storage limit, e.g. for small values of $F$, is to simply compute only parts of the spectrum in single runs. As we can see from Eqs.~(\ref{eq:05}) and (\ref{eq:25}), and also from the manifold structure of the eigenspectrum folded within the FZ (see sketch in Fig.~\ref{fig:00}), we can improve the performance by setting
various starting parameters $\tilde{\varepsilon}_j$ for the Lanczos algorithm, while defining $k_{\rm max}=k_{\rm min}\approx\Delta_r/2\omega_B$. 
We now can proceed by implementing $N+1$ independent Lanczos diagonalizations, each with its own sufficient number of Fourier components to promote one atom to 
an energy difference $\Delta_r/2$ starting from every $\tilde{\varepsilon}_j$. This only involves excitations until half way between two neighboring manifolds $M$ and $M\pm 1$. Thereby, we expect to obtain approximately a number $\sim d_M$ of eigenenergies around every $\tilde{\varepsilon}_j$. In this scheme, the dimension of the Krylov subspace ($n_p<d_s$) is much reduced along with the dimension of the $\mathcal M$. Consequently also the memory storage and the computational times are reduced for the individual runs.
In order to obtain the full spectrum around any chosen shift, e.g., $\tilde{\varepsilon}_0\approx 0$ as in Fig.~\ref{fig:00}, we wrap all eigenvalues (obtained around the others shifts $\tilde{\varepsilon}_{j\neq 0}\approx 0$) into the FZ defined by $\tilde{\varepsilon}_0$ as
\begin{eqnarray}\label{eq:27}
\varepsilon_j'\rightarrow\varepsilon_j-k_{0}\omega_B\,,\;\;{\rm with}\;\; 
k_0 = \left[\frac{\tilde{\varepsilon}_{j\neq 0}+\delta-\varepsilon_0}{\omega_B}\right]\in \mathcal Z\,.
\end{eqnarray}
Here $[\cdots]$ stands for the integer part function and $k_0$ is the number of Floquet zones needed
to map $\varepsilon_j'$s onto their corresponding Floquet eigenstates within the fundamental FZ centered at
$\tilde{\varepsilon}_0\approx0$. We defined $\delta \equiv \delta\varepsilon/2$. The folded eigenvalues thus
satisfy  $\varepsilon_j=\varepsilon_j{\rm mod}(\omega_B)$. As the final step we discriminate and eliminate copies of 
eigenstates using the fact that the overlap of physically different eigenvectors ideally vanishes. Our numerical 
orthogonalization criterion is that if $|\langle\varepsilon_i|\varepsilon_j\rangle|<10^{-7}$ we assume that the two vectors $|\varepsilon_i\rangle$ and $|\varepsilon_j\rangle$ ($i\neq j$) represent different eigenstates.

\begin{table*}[ht]\begin{center}
\caption{\label{tab:01} Comparison of the eigenenergies of the full spectrum (*) around $\tilde{\varepsilon}_0\approx 0$ (upper table) and
  $\tilde{\varepsilon}_2\approx 2\Delta_r$ (lower table) with folded eigenenergies with starting Lanczos parameters
  $\tilde{\varepsilon}_j=j\Delta_r \;\;(+\delta)$ ($^{\star,\diamond}$), with $\delta$ as defined after Eq.~(\ref{eq:27}). The
 Stark force is $F=\Delta\times 10^{-3}/\pi$, while the other parameters are the same as used in Fig.~\ref{fig:01-0}.}
\begin{tabular}{l|ccccc}\hline\hline
$\tilde{\varepsilon}_0$ &0.001349897362263* & 0.000531659341077*&-0.023753576110081*&-0.011655535562532*\\\hline
 &$\tilde{\varepsilon}_1{}^{\star},\;\tilde{\varepsilon}_1+\delta{}^{\diamond}$&$\tilde{\varepsilon}_2{}^{\star},\;\tilde{\varepsilon}_2+\delta{}^{\diamond}$&$\tilde{\varepsilon}_3{}^{\star},\;\tilde{\varepsilon}_3+\delta{}^{\diamond}$&$\tilde{\varepsilon}_4{}^{\star},\;\tilde{\varepsilon}_4+\delta{}^{\diamond}$\\
\cline{2-5}
 &0.001349897362262$^{\star}$&0.000531659341080$^{\star}$&-0.023753576110153$^{\star}$&-0.011655535570281$^{\star}$\\
 &0.001349897362263$^{\diamond}$&0.000531659341075$^{\diamond}$&-0.023753576110091$^{\diamond}$&-0.011655535561856$^{\diamond}$\\\hline\hline
$\tilde{\varepsilon}_2$ &5.040989227022449*&5.083166308393272*&5.071405551272533*&5.056877702765119*\\\hline
 &$\tilde{\varepsilon}_0{}^{\star},\;\tilde{\varepsilon}_0+\delta{}^{\diamond}$&$\tilde{\varepsilon}_1{}^{\star},\;\tilde{\varepsilon}_1+\delta{}^{\diamond}$&$\tilde{\varepsilon}_3{}^{\star},\;\tilde{\varepsilon}_3+\delta{}^{\diamond}$&$\tilde{\varepsilon}_4{}^{\star},\;\tilde{\varepsilon}_4+\delta{}^{\diamond}$\\
\cline{2-5}
 &5.040989227022449$^{\star}$&5.083166308393297$^{\star}$&5.071405551272531$^{\star}$&5.056877702765120$^{\star}$\\
 &5.040989227022451$^{\diamond}$&5.083166308392099$^{\diamond}$&5.071405551272880$^{\diamond}$&5.056877702764962$^{\diamond}$\\\hline 
\end{tabular}\end{center}
\end{table*}

In Table~\ref{tab:01}, we present an example of results based on the implementation with varying Lanczos shift parameters. We computed the full spectrum around
$\tilde{\varepsilon}_0\approx 0$ for $N/L=5/5$ with $d_s=402$, see the given eigenvalues marked by $({}^*)$. For this calculation, we used the parameters 
$\Delta_r\approx\Delta=2.53$, $2\pi F=2\Delta\times 10^{-3}$, and $W_b=0.025$. Therefore, the full diagonalization requires 
$k_{\rm max}\approx (N/2+0.125)\times101\approx 262$ (with fixed $k_{\rm min}=1$) Fourier components to obtain $d_s$ converged eigenvalues. 
The computation time for this computation was CPU time $\approx 20$ minutes. In comparison, we computed the subsets of eigenvalues around the shift parameters
$\tilde{\varepsilon}_{j>0}=\{\Delta_r,\cdots,N\Delta_r\}$, with $k_{\rm max}=k_{\rm min}\approx \Delta/2\omega_B=101/4\approx 25$ 
Fourier components. Then, by using Eq.~(\ref{eq:27}), we folded them into the FZ centered at $\tilde{\varepsilon}_0\approx 0$. 
For a single diagonalization process, the ${\rm CPU \ time} \approx 95$ seconds, that is, the total CPU time is $\approx 95\times(N+1)$ seconds $\approx 10$ minutes. 
This is at least by a factor of two faster than the time needed for one full diagonalization of the same problem. Hence, by using different shift parameters, a clear reduction of the computational time is expected for large system sizes. Then even computations for small tilts $F$ become possible. The resulting folded eigenvalues are shown in the remaining rows of Table~\ref{tab:01} (eigenvalues (${}^{\star,\diamond}$)) in comparison with the ones from the full spectrum around $\tilde{\varepsilon}_0\approx 0$ (${}^*$). 
Note that the agreement is excellent up to $12$ digits, at least for the first three eigenvalues belonging to the shift parameters $\{\tilde{\varepsilon}_{j=1,2,3}=j\Delta_r \;\;(+\delta)\}$.
There is a growing lack of accuracy as the energy distance increases between the FZ of interest ($\tilde{\varepsilon}_0$) and the one used for calculation of the spectrum increases ($\tilde{\varepsilon}_{j\neq 0}$). This problem can be partially corrected by choosing a different starting Lanczos parameter for the FZ of interest. That is, we redefine 
$\tilde{\varepsilon}_0\rightarrow\tilde{\varepsilon}_2=2\Delta_r$, and again compute the eigenvalues around $\{\tilde{\varepsilon}_{j\neq 2}\}$, and fold them using Eq.~(\ref{eq:27}). The result is shown in the lower part of the table. Here the agreement even with the largest eigenvalues, corresponding to those of the high lying manifolds, is much improved.
For our specific Wannier-Stark system, the trick of using different shift parameters is well supported by the quasi symmetry related to the occupation number $M$.
This symmetry is based on the fact that, for $F=0$ and small $W_x$ compared to the manifold band gap $\Delta_r$, the Hamiltonian can be approximated by the block form $\hat H=\oplus_{M=0}^{M=N}\hat H_M$. As explained in Sec.~\ref{sec:02}, this quasi symmetry remains approximately valid for $F\neq 0$, but only far from the resonance conditions $F=F_r$.

So far we have presented an efficient scheme for the numerical diagonalization of the Floquet Hamiltonian represented by the matrix $\mathcal M$. With both eigenvalues and eigenvectors at hand we shall now study the structure of the spectra with respect to changes of a control parameter, which is naturally the Stark force $F$ in our case. The extension and connection of these static properties of the spectra to the temporal evolution will then be discussed in section \ref{sec:05}.

\section{Detecting avoided crossings by eigenstate projection}
\label{sec:04}

After solving the Floquet eigensystem with the methods exposed in section~\ref{sec:03}, we now analyse the spectrum of our non-integrable many-body system.
The complexity in the spectra arises from the level-level repulsion of eigenstates as a function of, e.g., the control parameter $F$. Strong couplings in the resonant regime, in the vicinity of $F_r\approx \Delta/2\pi r$, typically lead to avoided crossings (ACs) in the parametric dependence of the levels. The occurrence of ACs with a wide distribution of crossing widths is associated with the onset of chaos in quantum systems (see e.g. the refs.~\cite{Haakebook,Carlitos01,Carlitos02,kol68PRE2003} in this context). The chaoticity of our system can be probed by means of statistical distributions, for instance, the nearest neighboring spacing distribution \cite{Haakebook,Wimb14} at fixed values of $F$. Another way to do this is by studying the distribution of ACs widths within a small range $\Delta F$ \cite{Kusmarek,Stoeckmannbook}. In the following, we focus on the latter approach as it characterizes a paradigm phenomenon which is directly connected to the eigenstates of our system.

The locations of ACs and their widths are difficult to detect numerically in large systems. Most methods are based on following the parametric changes of the curvature of
the energy levels \cite{PloetzLubach}, or on solving non-linear equations for the discriminant of the eigenvalue equation for a given Hamiltonian matrix \cite{RamanBhatta}.
The implementation of these methods affords the ordering of the energies and, for instance, the implementation of the discriminant method which is computationally expensive for large
system sizes. Here, we introduce an alternative method for detecting and characterizing ACs based on the statistical properties of the eigenvectors. The method is suitable for our Floquet system with periodic quasienergies since it avoids the implementation of routines for sorting and following the quasienergies. Our procedure may be applied to any kind of discrete quantum spectra making it an interesting method for the spectral analysis of generic quantum systems.

Relevant information on a complex quantum system can be extracted from the properties of the eigenvectors of the Hamiltonian under consideration \cite{TDittrich1991,PTWeidemueller}. 
Let us illustrate the idea with a simple example in our case. The eigenstates $|\varepsilon_j\rangle$ are connected 
to the TIF basis through a unitary transformation matrix as $\mathbf e=\mathcal U \mathbf s$,
with $\mathbf e^T=(\cdots,|\varepsilon_j\rangle,\cdots)$ and $\mathbf s^T=(\cdots,|s_{\alpha}\rangle,\cdots)$.
The transformation matrix elements are $(\mathcal U)_{\alpha,j}=\langle s_{\alpha}|\varepsilon_j\rangle$. 
In this way, since $\mathbf s=\mathcal U^{\dagger}\mathbf e$, any TIF vector can be written as
$|s_{\alpha}\rangle=\sum_j \langle \varepsilon_j|s_{\alpha}\rangle|\varepsilon_i\rangle$. Since
the occurrence of ACs is typically local in the spectrum, there are only a few eigenstates involved, two in the simplest case. Therefore, we may approximate
$|s_{\alpha}\rangle\approx \langle \varepsilon_j|s_{\alpha}\rangle|\varepsilon_j\rangle+\langle \varepsilon_{j+1}|s_{\alpha}\rangle|\varepsilon_{j+1}\rangle$,
where $|\langle \varepsilon_j|s_{\alpha}\rangle|^2\approx |\langle\varepsilon_{j+1}|s_{\alpha}\rangle|^2$. If now 
$|\langle \varepsilon_j|s_{\alpha}\rangle|^2+|\langle\varepsilon_{j+1}|s_{\alpha}\rangle|^2\approx
1$ is satisfied, the state $|s_{\alpha}\rangle$ approaches the hybridized state
\begin{eqnarray}\label{eq:28}
|s_{\alpha}^{(\pm)}\rangle\approx \frac{1}{\sqrt{2}}\left(|\varepsilon_j\rangle\pm|\varepsilon_{j+1}\rangle\right)\,.
\end{eqnarray}
Such a hybridization in the diabatic basis is typical of an AC. It effectively means that the effective number of participating states in the representation
of the local eigenstate is larger than one. Quantitatively, we can express this using the inverse participation ratio \cite{TDittrich1991}
\begin{eqnarray}\label{eq:29}
{\rm ipr}_{\alpha}(F) =\sum_j|\langle \varepsilon_j|s_{\alpha}\rangle|^4\,,
\end{eqnarray}
which is then also larger than one. Note that, for the state in Eq.~(\ref{eq:28}), we have $1/{\rm ipr}_{\alpha}=2$. 
Sufficiently far from an AC, ${\rm ipr}^{-1}_{\alpha}<2$, which implies that $1/{\rm ipr}_{\alpha}$ has a local maximum at the
exact position of the AC. The width of the AC is the energy separation $c \equiv \varepsilon_{i+1}-\varepsilon_{i}$. 
Note that this simple example does not involve the eigenenergies at all, but it does need the projections $p^{\alpha}_j=|\langle \varepsilon_j|s_{\alpha}\rangle|^2$. 

The method can be generalized as follows. Let $\mathbf p^{T}_{\alpha}$ be the vector of projections 
$\mathbf p^{T}_{\alpha}(F) = \left(\cdots,p_j^{\alpha}(F),\cdots\right)$. It represents the TIF state $|s_{\alpha}\rangle$ 
in the energy basis. Given two of these vectors, $\mathbf p$ and $\mathbf q$, they are never collinear unless they are the same, i.e., $\mathbf p\equiv\mathbf q$. 
Therefore, any change in the structure of $\mathbf p_{\alpha}$, as a function of the control parameter $F$, corresponds to the coupling to additional states.
This is exactly what happens at ACs. To quantify the structural transformation of $\mathbf p_{\alpha}$, a well suited measure is the 
fidelity function introduced in \cite{PloetzLubach} as
\begin{eqnarray}\label{eq:30}
G_{\alpha}(F,\delta F) &=&\mathbf p_{\alpha}^T(F)\mathbf p_{\alpha}(F+\delta F)\nonumber\\
                     &\approx&\mathbf p_{\alpha}^T(F)\left(1+\delta F\frac{\partial}{\partial F}\right)\mathbf p_{\alpha}(F)\,.
\end{eqnarray}
Here we have used a Taylor expansion up to the first order in $\delta F$:
\begin{eqnarray}\label{eq:31}
\mathbf p_{\alpha}(F+\delta F)=\mathbf p_{\alpha}(F)+\delta F \frac{\partial}{\partial F}\mathbf p_{\alpha}(F) + \cdots \,.
\end{eqnarray}
The differential operator $T_{\delta F} \equiv 1+\delta F\partial/\partial F$ acts as a translation operator in the $F$-coordinate. It obeys 
the composition property, i.e., $T_{\delta F_1}T_{\delta F_2} = T_{\delta F_1+\delta F_2}$, with $T_0=1$. 
Using the orthonormality condition, $\sum^{d_s}_{j=1}p^{\alpha}_j=1$, it is straightforwardly shown that
\begin{eqnarray}\label{eq:32}
\|\mathbf p_{\alpha}\|:=\sqrt{\mathbf p_{\alpha}^T\mathbf p_{\alpha}}=\sqrt{d_s\sigma^2+1/d_s} \,,
\end{eqnarray}
with $\sigma^2$ being the variance of $\mathbf p_{\alpha}$. Rewriting the fidelity function from Eq.(\ref{eq:31}), we obtain 
\begin{eqnarray}\label{eq:33}
G_{\alpha}(F,\delta F)= T_{\delta F/2}\sum_{j=1}^{d_s}(p_j^{\alpha})^2\,.
\end{eqnarray}
Here the last term, on the right-hand side is nothing but the inverse participation ratio 
${\rm ipr}_{\alpha}(F) =\sum\nolimits_j|\langle s_{\alpha}|\varepsilon_j(F)\rangle|^4$. Its inverse defines the 
effective number of participating states, just as in the example above for two states. Finally, we arrive at the relation
\begin{eqnarray}\label{eq:34}
\lim_{\delta F\rightarrow 0}G_{\alpha}(F,\delta F)={\rm ipr}_{\alpha}(F)=d_s\sigma^2+d_s^{-1}\,.
\end{eqnarray}
The latter implies that we have to find the maxima of the $1/{\rm ipr}$ in order to locate the position of the AC. 
The upper and lower bounds are $1/{\rm ipr}=1$ and $1/{\rm ipr_{\alpha}}=d_s$ corresponding to the vectors 
$\mathbf p^T_{\alpha}=\left(0,0,\cdots,1,\cdots,0,0\right)$ and $\mathbf p^T_{\alpha}=d_s^{-1}\left(1,1,\cdots,1,\cdots,1,1\right)$,  
respectively. This can be seen in Fig. 6, where the lower panel shows values between 1 and 2 for ACs with two coupled states, c.f. also the description after Eq. (29). 

\begin{figure}[t]
\centering 
\includegraphics[width=0.7\columnwidth]{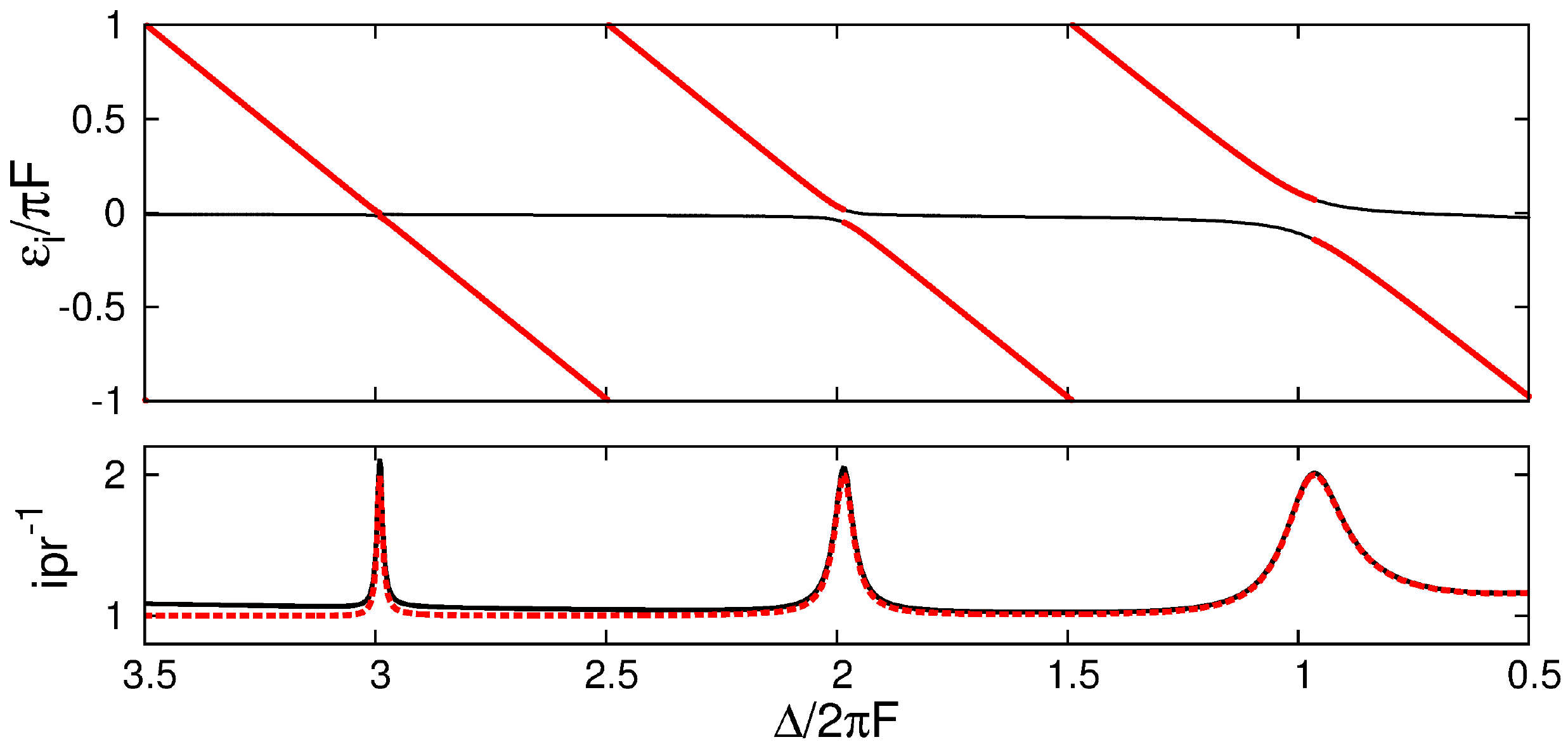}
\caption{\label{fig:02-0}(Color online) Single-particle spectrum (a) and the strong coupling at the ACs visible in the
  effective number of participating states $1/{\rm ipr}$ (b). The color discrimination 
  is as follows: eigenstates with $M_i\simeq 0$ (black lines) and eigenstates with $M_i\simeq 1$ (red/grey lines). 
	The band gap is $\Delta=2.53$.}
\end{figure}

\begin{figure}[t]
\centering 
\includegraphics[width=0.7\columnwidth]{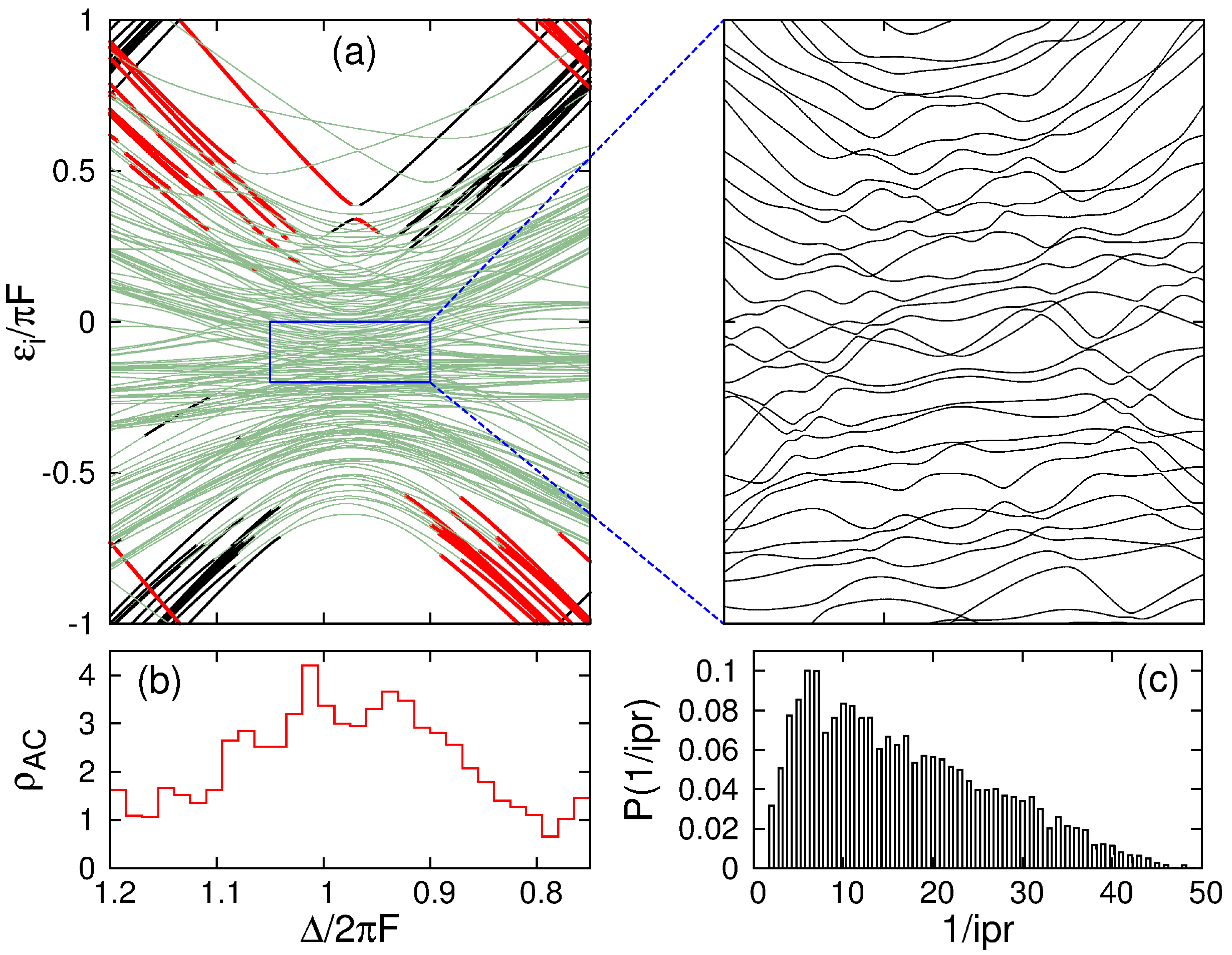}
\caption{\label{fig:02-1}(Color online) (a) Many-body spectrum around
  the first-order resonance, $r=1$, for $N/L=4/5$ and the $\kappa=0$ TIF subspace. 
  The zoomed region on the right highlights the clustering of avoided crossings in the center of the spectrum. The density of ACs is shown in (b).
  Panel (c) presents the normalized distribution of the effective number of participating states, $1/{\rm ipr}$, computed at all detected ACs within
  $\Delta F = 0.5 \times \Delta/2\pi$, i.e. in the full range of $F$ shown in (a). All panels (a-c) visualize the strong coupling, most prominent at the center of the spectrum.}
  The system parameters motivated in \cite{PhDThesis2013} are: $\Delta=0.56$, $J_a=0.06$, $J_b=-0.071$, $W_a=0.025$, $W_b=0.027$, $W_x=0.024$, $C_0=-0.095$, $C_{\pm 1}=0.04$, $C_{\pm 2}=0.004$.
\end{figure}

To estimate the width, $c$, of the detected AC, we appeal to the fact that the occurrence of an AC is typically local in energy space. Thereby, it is expected that at least two neighboring components of ${\mathbf p}_{\alpha}$ are maximal whenever the hybridization process happens. We can then study the functional behavior of the local energy difference 
$\omega \equiv c_j=\varepsilon_{j+1}-\varepsilon_{j}$ by means of the following projection vector distribution for two levels
\begin{eqnarray}\label{eq:35}
P^{(2)}_{\alpha}(\omega)&=&\sum_jp_j^{\alpha}p_{j+1}^{\alpha}\cdot\delta(\omega-(\varepsilon_{j+1}-\varepsilon_{j}))\nonumber\\
&=&\sum_j p_j^{\alpha}p_{j+1}^{\alpha}\cdot \lim_{\Gamma\rightarrow 0}\frac{1}{\pi}\frac{\Gamma}{(\omega-(\varepsilon_{j+1}-\varepsilon_{j}))^2+\Gamma^2}\,.
\end{eqnarray}
Here we expressed Dirac's delta in terms of a Lorentzian function with vanishing width $\Gamma$. $P^{(2)}_{\alpha}(\omega)$ as a single function efficiently detects the widths 
$\omega=\varepsilon_{j+1}-\varepsilon_{j}$ at which $P^{(2)}_{\alpha}(\omega)$ has a maximum. In a more general case, an AC can consist of
more than just two participating states. In this case, for a better estimation of the AC positions, we also use the following three level distribution 
$P^{(3)}_{\alpha}(\omega)=\sum_j p^{\alpha}_{j-1}p_j^{\alpha}p_{j+1}^{\alpha}\delta(\omega-c)$, with $c={\rm min}\{\varepsilon_{j+1}-\varepsilon_{j},\varepsilon_{j}-\varepsilon_{j-1}\}$. Because of the typically local feature of ACs in a discrete quantum spectrum (see the zoom in Fig. \ref{fig:02-1} as an example), it is practically sufficient to rely on the information content of both functions $P^{(2)}_{\alpha}$ and $P^{(3)}_{\alpha}$ for a reliable detection of ACs. We used this approach to find numerically the ACs, their positions, and their distributions within the resonant (RET) regime in an efficient and easy to automatize manner. 

In figure \ref{fig:02-1} we present the results for the detection ACs in our many-body Wannier-Stark system. In the zoomed region one can recognize the strong
level repulsion between the energy levels as a function of the force $\varepsilon_j(F)$. The ACs cluster in the region around $F_r$, as highlighted by the density of ACs 
$\rho_{AC}(F) = \#{\rm ACs}/d_s\Delta F$ in Fig.~\ref{fig:02-1}(b). Furthermore, we show the normalized distribution of the effective number of participating states 
in the entire resonance region $\Delta F$ in Fig.~\ref{fig:02-1}(c). We see the large number of effectively participating states in the resonant regime, arising from the
strong level mixing, in particular for the states with manifold number $0<M<N$ (see section \ref{sec:02}). 

\begin{figure}[t]
\centering 
\includegraphics[width=0.7\columnwidth]{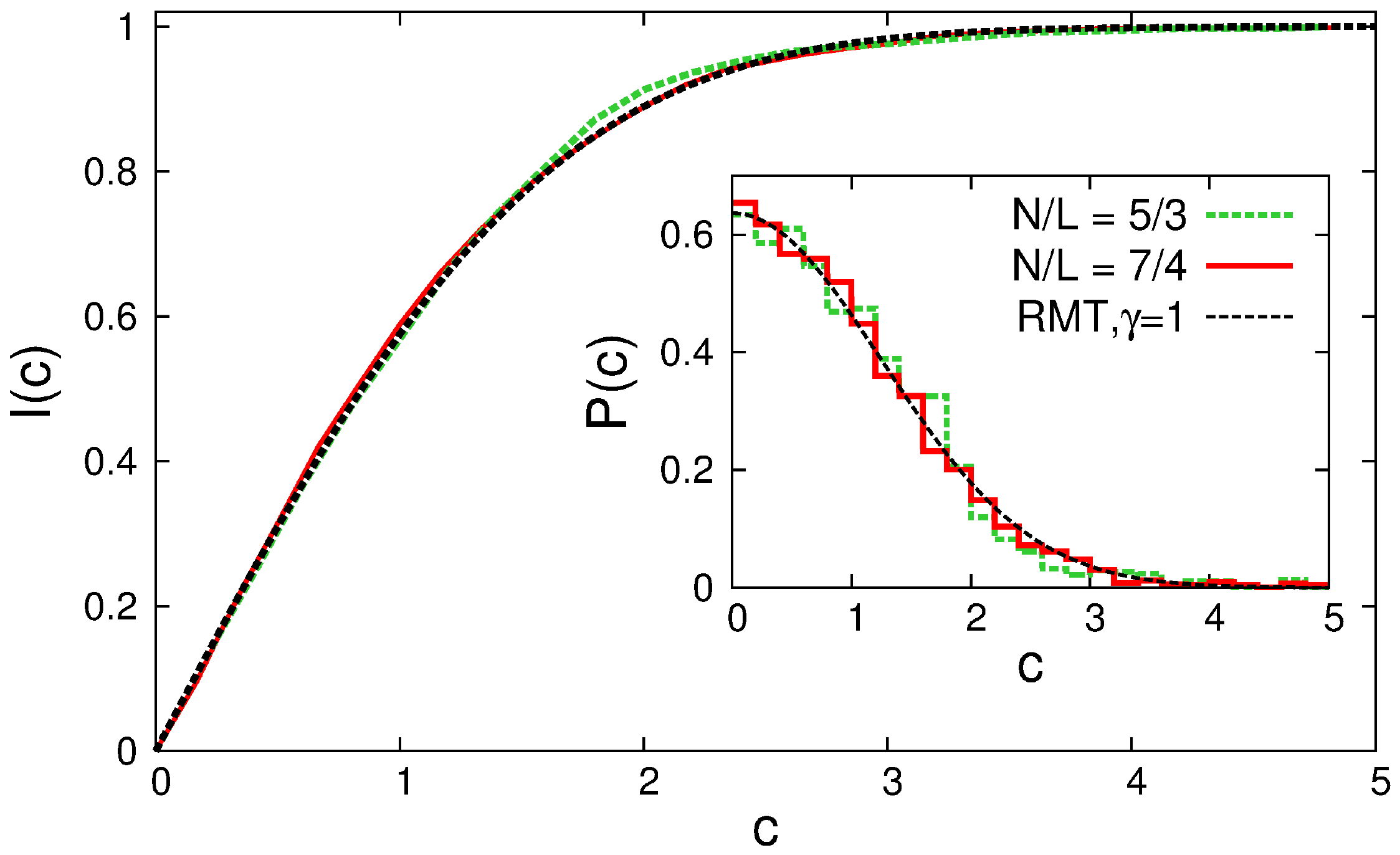}
\caption{\label{fig:03}(Color online) The inset shows the comparison between the theoretical predictions based on random matrices for $P(c)$ 
  (red/dark grey dashed line) and the numerically obtained distribution (black line). The main panel shows the respective cumulative distribution functions.
  Good agreement between the data and the random matrix theory for quantum chaos, expressed in the spectral statistics, clearly highlights 
  the strong coupling and, consequently, the complex dynamics in the temporal evolution of the system.
  Here $N/L=5/3$ ($d_s=84$) and $N/L=7/4$ ($d_s=858$) with a bandgap $\Delta = 0.155$ guaranteeing a large number of ACs ($\#{\rm ACs} = 4751$ for
  $N/L=5/3$ and $\#{\rm ACs} = 152145$ for $N/L=7/4$). $\Delta F$ is chosen as for the previous figure and a step size $\delta F = 10^{-4}$ was used for the computations.}
\end{figure}

As shown in ref.~\cite{Carlitos01} for our system, the number of ACs can be increased in our model by increasing the filling factor ($N/L>1$) and the interparticle interactions. This manifests itself  in a compression of levels within the Floquet zone \cite{Carlitos01}. Due to the lack of additional symmetries in the system, this leads to a crossover between regular and quantum chaotic spectra and dynamics, where the chaotic regime occurs for $N/L\sim 1$ and $\Delta \leq 1$ in the vicinity of the resonances at $F=F_r$. To characterize the quantum chaoticity, we need not only the pure relative occurrence frequencies of ACs, but also the statistics of their width distribution. We recall that our Hamiltonian in the $\kappa=0$ TIF subspace is represented by a real and symmetric Floquet matrix, which, within the context of random matrix theory (RMT), belongs to the so called circular orthogonal ensemble (COE) \cite{Haakebook}. Random matrix theory predicts for this case the following distribution of AC widths form \cite{Kusmarek}
\begin{eqnarray}\label{eq:36}
P(c) = (1-\gamma)\delta(c-c_0)+\frac{2\gamma^2}{\pi}\exp\left(-\frac{\gamma^2c^2}{\pi}\right)\,,
\end{eqnarray}
with normalized average distance $\langle c\rangle=1$. $\gamma$ is a fitting parameter. For $\gamma=0$, the formula characterizes a fully regular spectrum, with a relevant energy scale $c_0$. Conversely, $\gamma=1$ implies the randomization of the set of widths $c$, marking the completely chaotic regime (see Ref.~\cite{Kusmarek} for details). Here we confirm the expected chaoticity of our system by comparing our numerical distribution with the theoretical prediction for $N/L=5/3$ (green/light grey) and $N/L=7/4$ (red/dark grey) in Fig.~\ref{fig:03}. 
Interestingly, we see that even for a small system $N/L=5/3$ ($d_s=84$), the agreement between the chaotic distribution with $\gamma=1$ and the numerical 
one is still good. The agreement is excellent for the larger fillings $N/L=7/4$ ($d_s=858$), which guarantee better statistics. The number of detected ACs  is 4751 for
$N/L=5/3$ and is 152145 for $N/L=7/4$. The quality of the matching of the results of our numerical experiments with the theoretical predictions is best corroborated by computing the 
cumulative distributions $I(c)=\int_0^c P(c')dc'$, which are shown in the main panel of Fig.~\ref{fig:03}.

To summarize, we have shown that the detection and characterization of ACs is possible using the projection vectors $\mathbf p_{\alpha}$ and their
transformation as a function of a control parameter. The presented numerical method is well suited for any class of discrete eigenspectra, that is, not only for our Floquet spectra.
Since the ACs occurrence is mostly local in the spectrum, the use of the two and three-neighbor level distributions turned out to be sufficient
for well estimating the width of the ACs, such as shown in Fig.~\ref{fig:03}. The parametric evolution of the projection vectors studied here is inherently connected with
the dynamical evolution of the system. This will be seen in the next section after we will have discussed numerical techniques to evolve arbitrary initial states in time.

\section{Temporal evolution: algorithms and results}
\label{sec:05}

Having at hand the solution of the eigenvalue problem for our Floquet Hamiltonian in Eq. (\ref{eq:18}), we can easily propagate any given initial state $|\psi_0\rangle$ in time.
To do this, we best remind the reader on the representation of the time-evolution operator in the eigenbasis.  
With help of the closure relation $\sum_j|\varepsilon_j(t)\rangle\langle \varepsilon_j(t)|=\hat 1$, the evolution from time $t_1$ to time $t_2$ is obtained by applying the following
operator onto the initial vector
\begin{eqnarray}\label{eq:37}
 \hat{U}(t_2,t_1)&=&\sum_{j=1}^{d_s}e^{-i\varepsilon_j\Delta t} |\varepsilon_j(t_2)\rangle\langle \varepsilon_j(t_1)|\nonumber\\
                &=&\sum_{j=1}^{d_s}\sum^{k_{\rm max}}_{kk'=-k_{\rm  min}}e^{-i\varepsilon_j\Delta t}e^{i\omega_B kt_1} e^{-i\omega_Bk't_2}|
                \phi^{k'}_{\varepsilon_j}\rangle\langle \phi^{k}_{\varepsilon_j}|\,,
\end{eqnarray}
where $\Delta t=t_2-t_1$ \cite{BuchJosaB1995}. We can then compute any relevant physical quantity using this operator and the coefficients $c^0_{j}=\sum_{k}\langle \phi^{k}_{\varepsilon_j}|\psi_0\rangle$ 
defining the initial condition. Note that, in the case of $|\psi_0\rangle=|s_{\alpha}\rangle$, we have
\begin{eqnarray}\label{eq:38} 
c^{\alpha}_{j}=\sum_{k\alpha'}\langle s_{\alpha'}|(C^j_{k\alpha'})^*|\psi_0\rangle=\sum_{k}(C^j_{k\alpha})^*=\langle\varepsilon_i|s_{\alpha}\rangle\,. 
\end{eqnarray}
This straightforwardly shows us the direct connection between the initial condition for the time evolution and the projection vectors 
defined in the previous section. Starting the evolution at $t=0$, we can compute any observable described by a hermitian operator $\hat O$ using $\hat U(t,0)$ as follows
\begin{eqnarray}\label{eq:39}
O(t)=\langle\psi_t|\hat O|\psi_t\rangle
=\sum_{\alpha\beta}(\Lambda^0_{\alpha}(t))^*\Lambda^0_{\beta}(t)\langle s_{\alpha}|\hat O|s_{\beta}\rangle\,,
\end{eqnarray}
with 
\begin{eqnarray}\label{eq:40}
\Lambda^0_{\alpha}(t)=\sum_{jk}c_j^0e^{-i(\varepsilon_j+\omega_Bk)t}C^j_{k\alpha}\,.
\end{eqnarray}

Since for large systems, with increasing $N$ and $L$, the size of the Floquet matrices scales very unfavorably, i.e. exponentially with these parameters, we must ask ourselves whether it would not be better to use a direct propagation method for the given initial state.  We use an explicit propagator, e.g. the Runge-Kutta (RK) algorithm \cite{WPress}. More precisely, we resort to a fourth-order RK method with adaptive step-size based on a step-to-step error estimation. We applied an error threshold of $10^{-12}$ for the results shown here,
while no relevant changes are observed when comparing to thresholds between $10^{-9}\cdots 10^{-12}$. Very importantly, we replace the matrix-vector multiplication for the time integration by very efficient vector-vector multiplication. To do this, we store matrices in one-dimensional arrays and just cycle through the upper triangle of the matrix $\hat H'(t)$ in their order of appearance. The corresponding indexes are stored in vectors of integers associated with the non-zero matrix elements. The number of non-zero elements scales like $(\sim (8L+1)\times d_s)$ in our case, which saves a lot of allocation memory and considerably reduces the number of operations for the time integration. Since the RK algorithm does not intrinsically preserve the norm, we monitor the accuracy of the method via the preservation of the norm of the wave function along the time evolution. 

\begin{figure}[t]
\centering 
\includegraphics[width=0.65\columnwidth]{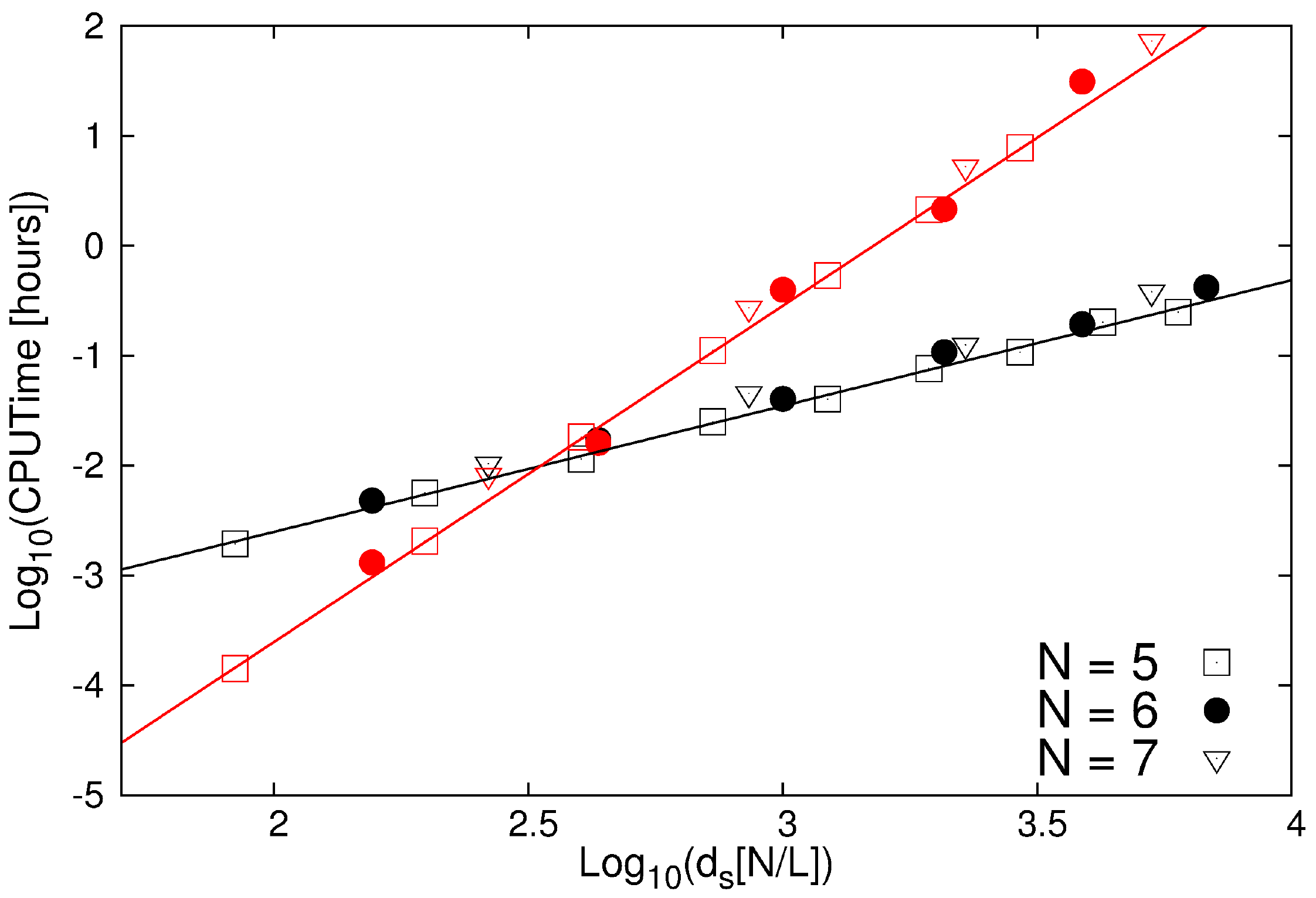}
\caption{\label{fig:06}(Color online) Polynomial growth of the CPU
  time with the size of the system $d_s$, for different atom numbers $N$ at fixed system sizes $L=5, 6, \ldots, 12$, respectively. 
  For both, the time evolution base on Eq. (\ref{eq:37}) and RK method the 
  CPU times grow as $\sim d_s^n$ with exponents
  $n=3.1\pm 0.1$ (red line for red/grey symbols) and
  $n=1.1\pm 0.1$ (black line for black symbols), respectively. 
  The parameters are the same as in Fig.~\ref{fig:01-0}.}
\end{figure}

We now compare the efficiency of both propagation methods by evolving in time the upper-band occupation number 
\begin{eqnarray}\label{eq:41}
M(t) = \langle \psi_0|\hat{U}^{\dagger}(t,0)\sum_{l=1}^L \hat n^b_l\hat{U}(t,0)|\psi_0\rangle = \sum_{l=1}^L  \langle \psi_0|\hat{U}^{\dagger}(t,0)\hat n^b_l\hat{U}(t,0)|\psi_0\rangle\,,
\end{eqnarray}
from $t=0$ to $t=1000\;T_B$ for a fixed Stark force $F$. The initial condition is given by the lowest energy state of the TIF basis
for the chosen filling factor $N/L$, for instance, states of the type $|\psi_0\rangle=|111\cdots\rangle\otimes|000\cdots\rangle$ at filling one. Figure \ref{fig:06} shows that the computation times scale algebraically with $d_s$, as defined around Eq. (\ref{eq:09}), for both methods, the Lanczos-based diagonalization with Eq. (\ref{eq:37}) and the RK propagation. The global scaling is much better for the RK method for large system sizes, here with exponent $n\simeq 1$, in contrast to the diagonalization with $n\simeq 3$. As expected, for not too large integration times, the RK method needs less CPU time, since here
the number of operations is much less than the ones needed for solving the eigenvalue problem. Moreover, the storage memory for the RK method is much less than the memory required in the Lanczos algorithm, which is already highly optimized with respect to memory requirements. But there is a crossover in efficiency measured by the CPU time. The absolute running times scale favorably for the Lanczos method only up to system sizes of $d_s \simeq 10^{5/2}$, while the RK integration is faster for large sizes $d_s > 10^{5/2}$, for the fixed overall integration time $t=1000\;T_B$ chosen in Fig. \ref{fig:06}. Only for evolutions up to very long times, the diagonalization will be favorable again. This is of relevance for small tilts $F$, since the characteristic time scale given by the Bloch period $T_B=1/F$, see also the discussion in section \ref{sec:03} around table \ref{tab:01}. Of course, the implementation of highly efficient matrix-vector multiplications, as stated above, is crucial for the given comparison. We finally note that the observed algebraic scalings of the CPU times are stable for a large range of system parameters, which is exemplified as a function of $N$ in Fig. \ref{fig:06}.

\begin{figure}[t]
\centering 
\includegraphics[width=0.6\columnwidth]{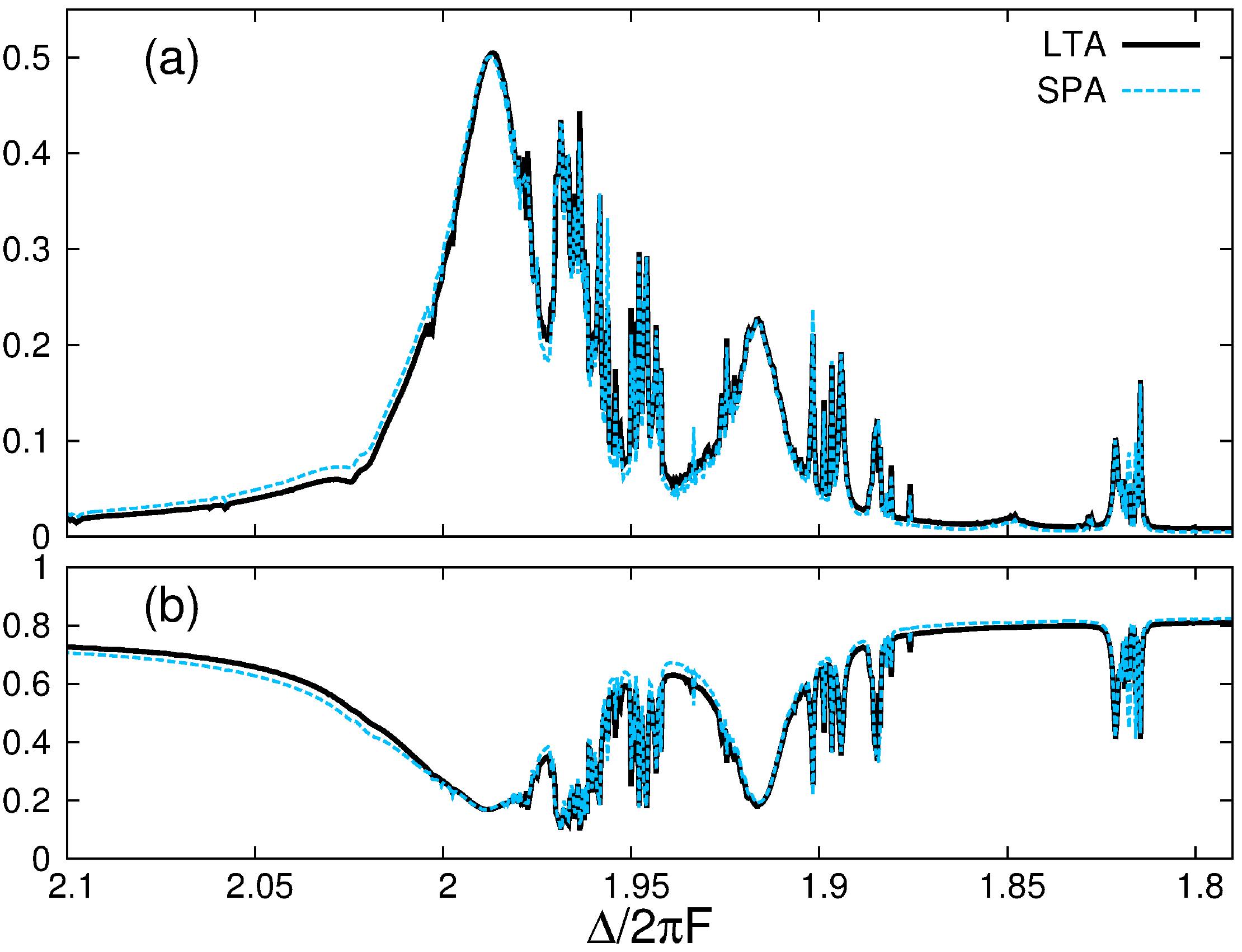}
\caption{\label{fig:07}(Color online) We show good agreement of the two methods detailed in the main text, the spectral (SPA) and long-time average (LTA)
  for (a) the upper-band occupation number and (b) the projector operator associated with the state $|s_{\alpha}\rangle=|111\cdots\rangle\otimes|000\cdots\rangle$. 
  The chosen filling is $N/L=5/5$, and we focus on the vicinity of the second
  order resonance with $r=2$. The evolution at every value of $F$ is done up to $t=10000\;T_B$.
  The other parameters are the same as applied in Fig.~\ref{fig:01-0}.}
\end{figure}

Coming back to our discussion in the previous section, we will see that the projection vectors defined there are intimately related with the temporal evolution of our system. For our study,
we now use the Lanczos method which turned out to be more efficient for large evolution times, for which the chaoticity is best observed in dynamical signals. The occurrence of avoided crossing is correlated with the strong, resonant coupling of at least two quantum states. The evolution of the vector $\mathbf p_{\alpha}$ can also be studied by means of the projector $\hat P_{\alpha}=|s_{\alpha}\rangle\langle s_{\alpha}|$.
Based on Eq.~(\ref{eq:37}), we compute the time evolution of the projector $\hat P_{\alpha}$, which reads
\begin{eqnarray}\label{eq:42}
P_{\alpha}(t)=\langle\psi_0|\hat{U}^{\dagger}(t,0)\hat P_{\alpha}\hat{U}(t,0)|\psi_0\rangle=|\Lambda^0_{\alpha}(t)|^2\,.
\end{eqnarray}
Here we set the initial condition to be $|\psi_0\rangle=|s_{\alpha}\rangle$. The system is now let to evolve up to some time $t$, and we compute the long-time average (LTA)
\begin{eqnarray}\label{eq:43}
\overline{P_{\alpha}(t)}&=&\lim_{\tau \rightarrow \infty}\frac{1}{\tau}\int^{\tau}_0\langle\psi_t|\hat  P_{\alpha}|\psi_t\rangle dt\nonumber\\
&=&\sum_{jk}|c_j^{\alpha}|^2|C^j_{k\alpha}|^2\nonumber\\
&+&\sum_{j\neq j'k\neq k'}c_j^{\alpha}(c_{j'}^{\alpha})^*C^j_{k'\alpha}(C^{j'}_{k'\alpha})^*\lim_{\tau\rightarrow\infty}\frac{1}{\tau}\int_{0}^{\tau}e^{-i(\omega_{jj'}-\omega_B\Delta_{k'k})t}dt \,,
\end{eqnarray}
with $\omega_{jj'}=\varepsilon_j-\varepsilon_{j'}$ and $\Delta_{k'k}=k'-k$. The integration over the entire time interval basically gives a Dirac delta function with argument 
$\omega_{jj'}-\omega_B\Delta_{k'k}$. The delta function will contribute only if $\omega_{jj'}=\Delta_{k'k}\omega_B$. Note that $\Delta_{k'k}$ is an integer, which implies that 
$\varepsilon_j=\varepsilon_{j'}+\Delta_{k'k}\omega_B$ maps the eigenenergy $\varepsilon_j$ from the fundamental Floquet zone onto a zone at a distance $\Delta_{k'k}$. The Floquet zones are equivalent, then without loss of generality $\varepsilon_j\rightarrow\varepsilon_{j'}$. Since degenerancies are excluded in our spectra due to just ACs in the chaotic regime, the second right-hand term in Eq.~(\ref{eq:43}) vanishes in the LTA, and we obtain
\begin{eqnarray}\label{eq:44}
\overline{P_{\alpha}(t)}\approx\sum_{jk}|c_j^{\alpha}|^2|C^j_{k\alpha}|^2\rightarrow \sum_j
p^{\alpha}_j\langle\varepsilon_j|\hat P_{\alpha}|\varepsilon_j\rangle={\rm ipr}_{\alpha}(F) \,.
\end{eqnarray}
This means that the system becomes ergodic during the evolution \cite{SpecErgo}. The effective number of participating states is hence large, and, in the extreme case,
the vector $\mathbf{p}^T_{\alpha}=\left(\cdots,p^{\alpha}_j,\cdots\right)$ approaches the equipartition condition 
$\mathbf p^T_{\alpha}=d_s^{-1}\left(1,1,\cdots,1,\cdots,1,1\right)$. We conclude that both, the appearance of ACs in the spectrum and the spectral averages (SPA) of
$\hat P_{\alpha}=|s_{\alpha}\rangle\langle s_{\alpha}|$ reflect themselves directly in the long-time behavior of the projection vectors. 

The LTA analysis can be straightforwardly extended to any other quantity, for instance the occupation number $\hat M$ of the upper energy band, whose long-time average reduces to 
$\overline{M(t)}\approx \sum_jp^{\alpha}_j\langle\varepsilon_j|\hat M|\varepsilon_j\rangle$. Figure \ref{fig:07} shows the comparison of the actual values and the ergodic limits of the observables mentioned above for $N/L=5/5$ ($d_s=402$), around a second order resonance for initial state $|\psi_0\rangle=|111\cdots\rangle\otimes|000\cdots\rangle$. In order to numerically compute $\overline{M(t)}$ and $\overline{P_{\alpha}(t)}$ we evolve the initial state up to time $t=10000 T_{B}$. Note that for both, $M(t)$ and $P_{\alpha}(t)$ their LTA and SPA correspond best close to ACs,
that is whenever $1/{\rm ipr}_{\alpha}$ is maximal. Finally, by evolving a given initial state $|\psi_0\rangle$ it is possible to detect the signatures of avoided crossings within the 
interesting spectral regions (here close to RET conditions). This gives an alternative and most likely easier route to the experimental measurements of signature of ACs and hence of quantum chaos in complex quantum systems, in the same spirit as the behavior of Bloch oscillation does in simpler single-band systems \cite{kolo03,Nagerl14}.


\section{Conclusions}
\label{sec:06}

In this paper we have presented exact numerical methods for the treatment of a one-dimensional Wannier-Stark system including strong interactions and two Bloch bands. This system has been of great experimental interest over the last decade \cite{greiner2009,Innsbruck2013,Wimb05,Mors06,Pisa2010,BOweak}, and experiments studying the strong coupling in the chaotic regime are just on their way \cite{Nagerl14}. For this implicitly time-dependent quantum system (since driven system by the Stark force $F$), we implemented two independent numerical approaches, one based on the combination of the Floquet theory and the Lanczos algorithm for symmetric matrices, and a second one based on a Runge-Kutta integration of a given initial state. In both cases, we ensured efficient matrix-vector multiplications and the memory storage of only non-zero elements in the corresponding Hamiltonian matrices. The computational times scale polynomially with the system size in both cases. Depending on the dimension of the system one or the other algorithm is more efficient for the temporal evolution over reasonable integration times. For large system sizes and not extremely long integration times, the direct and explicit Runge-Kutta integration proved to be more efficient, scaling with an exponent of approximately one with the system size. Nevertheless, the spectral analysis of the Floquet Hamiltonian is very useful for characterizing the dynamical properties. Interestingly, signatures of the chaoticity  of the system can also be found in the temporal evolution of appropriate observables. The avoided crossing scenarios in the spectrum as a function of the control parameter $F$ are directly related to the dynamical spreading across a given basis. This spreading even leads to complete ergodicity in the quantum chaotic regime of our model occurring at resonant tunneling conditions between the two energy bands. 

The numerical methods presented here are suitable for a larger variety of problems similar to the multi-band version of the Wannier-Stark system. For instance, 
periodic driving of bosonic many-body systems could be used to create highly entangled states \cite{lubasch}. Moreover, artificial gauge fields can be created by Stark forces or by time-periodic potentials \cite{gaugefields} in one or more spatial dimensions, possibly also in the strongly interacting regime with comparable resonant forces in the near future.Generally, as we showed for our case, the numerical performance may be improved exploiting the characteristic energy scales of the given system for an optimal choice of the numerical parameters. Our work shows that generic quantum lattice systems with strong many-body interactions should be tractable by exact methods up to Hilbert space sites of about $10^5$ on standard scientific workstations. Using sophisticated parallelization techniques and larger computing clusters with access to much more memory, one might gain at least another order of magnitude for the maximally treatable Hilbert space sizes \cite{Krug}.

\section{\bf Acknowledgments}
We kindly acknowledge contacts and discussions within the COST Action MP1209 ``Thermodynamics in the quantum regime". Furthermore, it is our great pleasure to thank Dominique Delande and Andreas Buchleitner for the central ingredients of the Floquet-Lanczos approach used here.


\end{document}